\documentclass[aps,pra,superscriptaddress,twocolumn]{revtex4-1}

\usepackage{gensymb}
\usepackage{amsmath}
\usepackage{amssymb}
\usepackage{bm} 
\usepackage{color}
\usepackage[linkcolor=blue, colorlinks=true, urlcolor=blue,
citecolor=red]{hyperref}
\usepackage{graphicx}
\usepackage[caption=false]{subfig}
\usepackage{bbold}
\usepackage{verbatim}
\usepackage{appendix}
\usepackage{amsthm}
\usepackage[normalem]{ulem}
\usepackage{framed}

\theoremstyle{plain}

\newcommand{\ket}[1]{\left| #1 \right>} 
\newcommand{\bra}[1]{\left< #1 \right|} 
\newcommand{\ketbra}[2]{\ket{#1}\hspace*{-0.mm}\bra{#2}}

\newcommand{\Id}{\mathbb{1}}

\newcommand{\Tr}[1]{\mathrm{\text{Tr}}\left[#1\right]}
\newcommand{\TrS}[2]{\mathrm{\text{Tr}}_{#2}\left[#1\right]}
\newcommand{\aop}{\hat{a}}
\newcommand{\bop}{\hat{b}}
\newcommand{\rop}{{\bf \hat{r}}}
\newcommand{\xop}{\hat{x}}
\newcommand{\pop}{\hat{p}}

\newcommand{\bxi}{{\pmb \xi}}

\newcommand{\bz}{{\pmb 0}}

\newcommand{\Eq}[1]{Eq.~\eqref{#1}}
\newcommand{\minus}{\! - \!}
\newcommand{\plus}{\! + \!}
\newcommand{\eq}{\! = \!}
 
\begin{document}

\title{Versatile Gaussian probes for squeezing estimation}

\author{Luca Rigovacca}
\affiliation{Blackett Laboratory, Imperial College London, London SW7 2AZ, United Kingdom}   

\author{Alessandro Farace}
\affiliation{Max-Planck-Institut f¨ur Quantenoptik, Hans-Kopfermann-Straße 1, 85748 Garching, Germany}  

\author{Leonardo A. M. Souza}
\affiliation{Universidade Federal de Vi\c{c}osa - Campus Florestal, LMG818 Km6, Minas Gerais, Florestal 35690-000, Brazil}
\affiliation{Centre for the Mathematics and Theoretical Physics of Quantum Non-Equilibrium Systems (CQNE), School of Mathematical Sciences, The University of Nottingham, University Park, Nottingham NG7 2RD, United Kingdomm} 

\author{Antonella De Pasquale}
\affiliation{NEST, Scuola Normale Superiore and Istituto Nanoscienze-CNR, I-56126 Pisa, Italy}  

\author{Vittorio Giovannetti}
\affiliation{NEST, Scuola Normale Superiore and Istituto Nanoscienze-CNR, I-56126 Pisa, Italy}  

\author{Gerardo Adesso}
\affiliation{Centre for the Mathematics and Theoretical Physics of Quantum Non-Equilibrium Systems (CQNE), School of Mathematical Sciences, The University of Nottingham, University Park, Nottingham NG7 2RD, United Kingdomm}

\begin{abstract}
We consider an instance of ``black-box'' quantum metrology in the Gaussian framework, where we aim to estimate the amount of squeezing applied on an input probe, without previous knowledge on the phase of the applied squeezing. By taking the quantum Fisher information (QFI) as the figure of merit, we evaluate its average and variance with respect to this phase in order to identify probe states that yield good precision for many different squeezing directions. We first consider the case of single-mode Gaussian probes with the same energy, and find that pure squeezed states maximize the average quantum Fisher information (AvQFI) at the cost of a performance that oscillates strongly as the squeezing direction is changed. Although the variance can be brought to zero by correlating the probing system with a reference mode, the maximum AvQFI cannot be increased in the same way.
A different scenario opens if one takes into account the effects of photon losses: coherent states represent the optimal single-mode choice when losses exceed a certain threshold and, moreover, correlated probes can now yield larger AvQFI values than all single-mode states, on top of having zero variance.
\end{abstract}

\maketitle

\section{Introduction}\label{sec: Intro}
Quantum metrology is one of the most developed fields within the framework of quantum information, and it has been the object of several studies in the last decade \cite{Braunstein&Caves_1994,Giovannetti_2006,Giovannetti_2011,Dobrzanski_2012,Toth2014a,Dobrzanski_2014,Dobrzanski_2015}. As a matter of fact, precise measurements and observations are a fundamental part of the scientific method and the possibility of exploiting quantum mechanics in order to improve the estimation precision over the best classical strategy is thus very appealing. Quantum metrology finds application in a wide range of situations where one is interested in gaining precise information about a system or a physical process \cite{Schnabel_2010,Dobrzanski_2013,Taylor_2014,Matthews_2016,Jachura_2016}.
 
In order to gather information about a certain system, the most common approach consists of sending a known probing system to it, collect the probe after it had the chance to interact with the target system, and look for differences with respect to the original state that was sent in. From this comparison, our knowledge about the state of the target system and/or the physical processes taking place during the interaction can be improved.
In analogy with this simple schematic picture, a typical estimation protocol can be divided into three steps: (i) probe preparation, (ii) interaction between probe and system of interest, and (iii) readout measurement on the evolved state of the probe.
These three stages are not independent from each other, and the quality of the overall estimation depends on how well they work together. For example, when the probe is being prepared one should make sure it can be significantly altered by the process under investigation. Similarly, any change in the probe is not useful at all if it cannot be detected by the performed measurement. Here, as in most theoretical studies, we focus on the interplay between the first and the second stage mentioned above. We assume that any measurement allowed by the laws of quantum mechanics can be performed, and we investigate how to tailor the choice of the probe depending on the information available at the first stage.

In the simplest scenario, the evolution of the probe is considered to be completely known in advance, with the exception of a real parameter that has to be estimated. A typical example is that of a unitary evolution of the probe, in which the generating Hamiltonian is known up to a multiplicative parameter (e.g. this can be the case of a probe interacting with an external classical field of unknown strength that we want to estimate). In this case, the optimal performance can be achieved by choosing a probe which is maximally coherent with respect to the eigenbasis of the Hamiltonian \cite{Giovannetti_2011}. However, there are many situations where a larger degree of ignorance limits the optimization of the probe. For example, in the presence of noise the actual evolution of the probe has an intrinsic degree of randomness which can strongly affect the estimation precision. In this case a good strategy is to look for states of the probe that guarantee a certain level of precision for any possible realization of the noise. 
We can discuss this situation on a more abstract level by formulating a game in which we are given a set of possible probe-system interaction processes, and we are asked to prepare many identical copies of a probe that will interact with the system in order to gain information about some parameter. However, at this point we still do not know which specific interaction will be realized. Then, one interaction is chosen at random and communicated to us. At last, we can let the probes sequentially interact with the system, and make an optimal measurement on the evolution of each probe in order to estimate the parameter. This scenario goes under the name of ``black-box'' metrology \cite{Girolami_2014}. A different situation could arise if the interaction is free to fluctuate from one probe to the next, without our knowledge \cite{Nichols_16} or such that each choice is communicated to us upon recollection of the probes.
Depending on the considered scenario, and on the set of the allowed encoding transformations, it is natural to wonder which input probes lead to a good estimation precision over the whole set of possible interactions, and which are the resources responsible for this behavior.

A scenario of black-box metrology has been recently studied for unitary encodings in finite dimensional systems, where the eigenvalues of the generating Hamiltonian are fixed and known, while the exact eigenbasis is left unspecified \cite{Girolami_2014,Farace_2014,Girolami_2013,Roga_2014}.
One way to ask for versatility is to maximize the worst-case estimation precision over the set of possible encodings (or equivalently over the set of isospectral Hamiltonians). In this case it has been found that the input probe needs to be correlated with another ancillary system, which is kept as a reference and measured together with the probe at the measurement stage. This is because any local probe undergoing a unitary evolution is left unchanged by a Hamiltonian diagonal in its same eigenbasis and the worst-case precision becomes trivially zero. Although entangled states guarantee the highest minimal precision, interestingly the presence of entanglement is not a necessary condition in order to obtain a nonzero worst-case performance \cite{Adesso_2016,Roga_2016,Bromley2016a,Braun2017a}. Indeed, the key resource in this and similar discrimination tasks is a weaker form of correlation known as ``quantum discord'' \cite{Ollivier_2001,Henderson_2001}. A complementary approach to versatility consists of looking for the probe that guarantees the best \emph{average} (instead of minimal) performance. In Ref. \cite{Farace_2016} it is shown that from this perspective the presence of correlations is helpful but does not represent the only important parameter, as local purity also plays a fundamental role.

When continuous variable systems are concerned, one typically considers additional realistic constraints on the probe (such as finite energy, finite correlations) which give access only to a limited portion of the Hilbert space and make the analysis more involved. Up to now only the former of the two aforementioned approaches to black-box metrology (i.e., guaranteeing minimal performances) has been investigated in the framework of Gaussian states and operations \cite{Ferraro_2005,Weedbrook_2012,Adesso_2014_GaussReview}, and quantum discord has once again been identified as the important figure of merit for these discrimination tasks \cite{Adesso_2014_GIP, Rigovacca_2015, Roga_2015}.  

In this work, we move forward along two different directions. On one hand, we take the perspective of looking at the average performance associated with a certain input Gaussian probe. On the other hand, in the second half of this paper we also study the effects of a noisy encoding operation, moving away from the most common unitary setting. It is worth stressing that in all previous studies of Gaussian black-box metrology, the set of possible encoding Hamiltonians has been obtained by applying a generic Gaussian unitary operation to a harmonic Hamiltonian \cite{Adesso_2014_GIP, Rigovacca_2015, Roga_2015}. Although this choice represents the Gaussian equivalent of a generic change of basis in the finite-dimensional domain, it has the disadvantage of introducing energy into the probe, at random and for free. This is because a generic Gaussian unitary operation includes the application of squeezing. Intuitively, adding energy to the probe increases the estimation precision. While this does not affect the study of the worst-case precision, it can instead arbitrarily increase the average precision. Therefore, to make our model meaningful, we will apply only ``passive'' Gaussian unitary operations (e.g., optically obtainable via beamsplitters and phase shifters) to a fixed seed Hamiltonian. In the following we will choose the seed Hamiltonian to be a single-mode squeezing Hamiltonian, and passive Gaussian unitaries are then identified by all single-mode phase rotations.
  
The problem of estimating the parameter of a squeezing Hamiltonian has been investigated in the past, by looking at its effect on the Hilbert space of the radiation field \cite{Milburn_1994,Chiribella_2006}, or by using Gaussian \cite{Paris_09,Safranek_2015,Safranek_16_OptGauss} or non-Gaussian probes obtained via Kerr interactions \cite{Genoni_2009}.  
The goal of this paper, instead, is to understand which Gaussian probe yields the optimal average performance for the task of estimating the amount of squeezing applied on the probing system by an external device, without prior information on the direction of application. This direction could be fixed, but initially unknown, or could randomly fluctuate from one encoding operation to another. If the state of the probe that will be used in all experiments is fixed beforehand, our results indifferently apply to both these scenarios, as long as full information on the direction is available at the measurement stage. 
In particular, we want to discuss whether the presence of input correlations can lead to an improvement over the optimal single-mode result. 
We start by considering a noiseless setup, and we characterize the single-mode states that yield the best average estimation precisions. We compare their performance with the precision obtainable by sending half of a two-mode squeezed state to the squeezing device, while keeping the other half as reference. Although the average estimation precision reached by this paradigmatic class of correlated bipartite states equals that of the optimal single-mode probes, we can show that the correlated probes have the advantage of a stable performance over all squeezing direction, at the cost of introducing extra photons for the reference beam. The advantage of using correlations will become even more important when noise is added to the process, in the form of photon losses during the transmission of the probe signal to and from the squeezing device. Indeed, a numerical analysis reveals how in this case the presence of correlations can even improve the average precision over the value associated with the optimal single-mode probe.

The following sections are organized as follows. After presenting some preliminary notions in Sec.~\ref{sec: preliminaries},  in Sec. \ref{sec: model} we formally introduce the black-box metrology model we are considering, and show a physical situation where it could arise. We analytically solve the problem in absence of noise in Sec. \ref{sec: noiseless}, while in Sec. \ref{sec: noisy} we numerically study the same situation in presence of losses. We present our conclusions in Sec. \ref{sec: Conclusions}, and further technical comments can be found in the appendices.

\section{Preliminary Notions} \label{sec: preliminaries}
In order to assess the performance of a given probe in an estimation task, in the following we will use the quantum Fisher information (QFI) \cite{Paris_2009}. In this section we provide its definition, and we introduce the basic formalism used to describe Gaussian states of continuous variable systems \cite{Ferraro_2005,Weedbrook_2012,Adesso_2014_GaussReview}. No original contribution will be presented here, with the only exception of \Eq{eq: Formal singMod QFI}: although formally equivalent to the result of Pinel and coworkers \cite{Pinel_2013}, the use of this formula for the QFI of single-mode Gaussian states will simplify the calculations in the example studied in this paper. 

\subsection{Estimation theory and quantum Fisher information}\label{sec: estimation theory}
The problem of estimating a parameter characterizing a certain evolution by repeatedly measuring its output has a long history and was originally studied in a classical framework. The typical situation here involves a parameter-dependent probability distribution $p(x|\epsilon)$. The goal is to obtain the best possible estimation of the parameter $\epsilon$ by sampling many times the random variable $x$ distributed according to $p(x|\epsilon)$. A well known result in classical estimation theory, which goes under the name of Cram\'{e}r-Rao bound \cite{CramerBook}, states that the Root Mean Square Error (RMSE)  of any unbiased estimator $\hat{\epsilon}$ of the parameter $\epsilon$ has to satisfy the following inequality:
\begin{equation}\label{eq: classical CRB}
\delta {\hat{\epsilon}}  \geq \frac{1}{\sqrt{M F_{\epsilon}}}.
\end{equation}
On the right-hand side, $M$ represents the number of performed samplings, and $F_{\epsilon}$ is the Fisher information of the process, defined by
\begin{equation}\label{def: Fisher info}
	F_\epsilon = \int \text{d}x \, p(x|\epsilon) \left[\partial_\epsilon \ln p(x|\epsilon)\right]^2. 
\end{equation}
Importantly, the bound in \Eq{eq: classical CRB} can be saturated in the asymptotic limit of many measurements, for example if the estimator $\hat{\epsilon}$ is obtained by maximizing the likelihood of the recorded events.

In the quantum counterparts of the aforementioned situation, the parameter $\epsilon$ is encoded in a quantum state $\rho_\epsilon$, typically obtained by applying a completely positive and trace-preserving (CPT) map $\Phi_\epsilon$ to a known input probe $\rho$ \cite{Nielsen&Chuang}, so that:
 \begin{equation}\label{eq: generic encoding}
\rho_\epsilon = \Phi_\epsilon[\rho].
\end{equation}
In order to obtain information about the encoding device, and thus on $\epsilon$, a generic positive operator valued measurement (POVM) has to be performed on $\rho_\epsilon$. This kind of measurement is characterized by a set of positive operators $\{E_x\}_x$, satisfying $\sum_x E_x = \Id$. Once applied on the encoded states, it yields the measured value $x$ with probability
\begin{equation}
p|_{\{E_x\}_x}(x|\epsilon) = \Tr{\rho_\epsilon E_x},
\end{equation}
from which the associated classical Fisher information $F_\epsilon|_{\{E_x\}_x}$ can be evaluated via \Eq{def: Fisher info}.
Clearly, the ultimate precision allowed by quantum mechanics for the unbiased estimation of $\epsilon$ is obtained by optimizing over all POVMs. In this way, it is possible to obtain the quantum Cram\'{e}r-Rao bound \cite{Paris_2009}, which states
\begin{equation}
\label{eq: quantum CRB}
\delta {\hat{\epsilon}}  \geq \frac{1}{\sqrt{M H_\epsilon(\rho)}},
\end{equation}
where $H_\epsilon$ is the quantum Fisher information (QFI) associated with the encoded state $\rho_\epsilon$, obtained from $\rho$ as in \Eq{eq: generic encoding}. This quantity is defined as
\begin{equation}\label{def: QFI}
H_\epsilon[\rho] = \Tr{\rho_\epsilon L_\epsilon^2},
\end{equation}
where $L_\epsilon$ is a Hermitian operator which goes under the name of symmetric logarithmic derivative (SLD) and satisfies the relation
\begin{equation}
\rho_\epsilon L_\epsilon + L_\epsilon \rho_\epsilon = 2 \,\partial_\epsilon \rho_\epsilon.
\end{equation}
Alternatively, it has been shown that the QFI is closely related to the second-order expansion of the Bures distance \cite{Bures_1969}, or equivalently of the Uhlmann fidelity ${\mathcal F (\rho_1,\rho_2)= \left(\Tr{\sqrt{\sqrt{\rho_1}\rho_2\sqrt{\rho_1}}}\right)^2}$ \cite{Uhlmann_1976}:
\begin{equation}\label{def: QFI as Bures expansion}
H_\epsilon[\rho] = 8 \lim_{\text{d}\epsilon\to 0} \frac{1-\sqrt{\mathcal F(\rho_\epsilon,\rho_{\epsilon+\text{d}\epsilon})}}{\text{d}\epsilon^2}.
\end{equation}
Rigorously speaking, this last equality is true as long as the rank of $\rho_{\epsilon}$ does not change in correspondence of the value of $\epsilon$ in which \Eq{def: QFI as Bures expansion} is calculated. If this should not be the case, Ref. \cite{Safranek_2016} recently showed that  
\Eq{def: QFI as Bures expansion} should be corrected by adding a term that involves second derivatives of the vanishing eigenvalues. As a consequence, the QFI might become discontinuous at those pathological points. 
Finally, if one is interested in obtaining the ultimate precision for the estimation of the parameter characterizing the CPT encoding map $\Phi_\epsilon$, an optimization has to be performed over the probe $\rho$.

We conclude this overview of quantum estimation theory with a few remarks. At first, let us point out that the projective measurement on the eigenbasis of the SLD operator $L_\epsilon$ has always a Fisher information equal to the QFI \cite{Paris_2009}. This assures that the bound of \Eq{eq: quantum CRB} is \emph{a priori} tight, even though the necessary control needed to perform this particular measurement is often out of experimental reach. Even in this situation, however, the QFI can be considered a figure of merit for the probe state, which has the potential to be very susceptible to small changes of the parameter $\epsilon$ that characterizes the encoding channel. 
As a second remark, note that in general the QFI depends on the real parameter $\epsilon$. This is why the QFI gives the ultimate precision attainable in a \emph{local} estimation: typically one already has some knowledge of $\epsilon$, and is interested in finding small fluctuation around this approximately known value . The situation simplifies for unitary encodings, where $\Phi_\epsilon$ can be actually written as $U_\epsilon \rho U_\epsilon^\dagger$, with $U_\epsilon U_\epsilon^\dagger = U_\epsilon^\dagger U_\epsilon = \Id$ and $\epsilon$ represents a global phase \cite{paramestimPRA}. In this case, $H_\epsilon(\rho)$ is independent from $\epsilon$, and therefore so is the ultimate precision attainable by $\rho$.

\subsection{Gaussian states of continuous variable systems}
In this paper we consider continuous variable systems composed by one or two bosonic modes described by the annihilation operators $\aop$ and $\bop$, which satisfy the canonical commutation relations
\begin{equation}	
[\aop,\aop] = [\bop,\bop] = [\aop,\bop] = [\aop,\bop^\dagger] = 0, 
\end{equation}
\begin{equation}
[\aop,\aop^\dagger]=[\bop,\bop^\dagger] = 1.
\end{equation}
In the following we will label with $A$ and $B$ the systems associated with the modes described respectively by $\aop$ and $\bop$.
The associated quadratures ${ \xop_A = (\aop^\dag + \aop)/\sqrt{2} }$, ${ \pop_A = i (\aop^\dag - \aop)/\sqrt{2} }$, and similarly $\xop_B$ and $\pop_B$, can be combined to form the vector $\rop = (\xop_A,\pop_A,\xop_B,\pop_B)^\intercal$. This appears in the definition of the covariance matrix associated with any state $\rho$ of this continuous variable system:
\begin{equation}
\Gamma = \Tr{\rho\,\{\rop - \bxi,\rop^\intercal - \bxi^\intercal\}_+},
\end{equation}
where $\{\cdot,\cdot\}_+$ represents the anti-commutator and ${\bxi=\Tr{\rho\, \rop}}$ is the associated displacement vector. The Robertson-Schr\"{o}dinger uncertainty relation, which all physical states must satisfy, in this language can be written as
\begin{equation} \label{def: RS relation}
\Gamma + i \Omega \geq \bz,
\end{equation}
where $\Omega$ is the standard symplectic form
\begin{equation}
\Omega = \bigoplus_j \left(\begin{array}{cc}
0 & 1\\
-1 & 0
\end{array}\right).
\end{equation}
Here, and whenever not explicitly mentioned, the sum over $j$ runs over the two bosonic modes of the system.

In this paper we will focus on Gaussian states, defined as those density matrices $\rho$
characterized by a characteristic function $\chi_\rho({\bf z}) = \Tr{\rho\, e^{- i {\bf z}\Omega \rop}}$, with ${\bf z} \in \mathbb R^2$, that is the inverse Fourier transform of a Gaussian function. They are completely characterized by the first and second moments of the latter, i.e., by the vector $\bxi$ and the matrix $\Gamma$ previously defined. 	
At this point it is useful to mention Williamson's theorem, which allows us to decompose every covariance matrix $\Gamma$ into the form
\begin{equation}\label{def: Williamson decomposition}
\Gamma = T D T^\intercal.
\end{equation} 
Here, $T$ is a matrix of the real symplectic group $Sp$, i.e., such that $T \Omega T^\intercal = \Omega$, and $D = \bigoplus_j \nu_j \Id_2$ is a block-diagonal matrix whose blocks are multiples of the $2\times 2$ identity matrix. These proportionality coefficients $\{\nu_j\}_j$ are called ``symplectic eigenvalues'' of the Gaussian states, are constrained by \Eq{def: RS relation} to be larger than or equal to 1, and can be found as regular eigenvalues of the positive-definite matrix $|i\Omega\Gamma|$. A pure Gaussian state is characterized by $\nu_j \equiv 1$, and we refer to a state with $T = \Id$ as to a ``thermal state''.

A unitary evolution $\mathcal U$ that maps the set of Gaussian states into itself can be fully characterized by its action on $\rop$:
\begin{equation}\label{eq: symplectic action}
  \mathcal U[\rop] =   U^{-1} \rop + \bxi^{(\mathcal U)},
\end{equation}
where $U \in Sp$ and $\bxi^{(\mathcal U)}$ is a real vector. The particular Gaussian unitary evolutions that will be considered in this work are characterized by ${\bxi^{(\mathcal U)} = \bz }$. In this case, the covariance matrix and the displacement vector associated with $\mathcal U[\rho]$ can be written in terms of $\bxi$ and $\Gamma$, which characterize the input Gaussian state $\rho$, via the following mapping:
\begin{equation}\label{eq: CV unitary change}
\Gamma_{} \stackrel{\mathcal U}{\longrightarrow} U \Gamma U^\intercal, \qquad \bxi \stackrel{\mathcal U}{\longrightarrow} U\bxi.
\end{equation}
In particular, we will focus on single-mode phase rotations and squeezing operations, that take an input state $\rho_A$ respectively to $\mathcal R_{\theta}^{(A)}[\rho_A] = e^{-i\theta \aop^\dagger \aop} \rho_A e^{+i\theta \aop^\dagger \aop}$ and 
 $\mathcal S^{(A)}_{\alpha}[\rho_A] = e^{-\frac{\alpha}{2}\left(\aop^{\dagger 2} - \aop^2\right)}\rho_A e^{+\frac{\alpha}{2}\left(\aop^{\dagger 2} - \aop^2\right)}$. These Gaussian unitary maps can be characterized by the symplectic matrices:
\begin{equation}\label{def: symplectic matrices}
R_\theta^{(A)} \eq \left(\begin{array}{cc}
\cos\theta & \sin \theta\\
\minus \sin\theta & \cos \theta
\end{array}\right), \quad 
S_\alpha^{(A)} \eq \left(\begin{array}{cc}
e^\alpha & 0\\
0 & e^{-\alpha}
\end{array}\right),
\end{equation}
which take the role of $U$ in \Eq{eq: symplectic action} and \eqref{eq: CV unitary change}.
The first of these operations simply rotates the quadratures of the state in phase space, while the effect of squeezing is to reduce the variance of one quadrature while increasing the other.
The CPT maps that preserve the Gaussian character of a state are not limited to the set of Gaussian unitary evolutions, but include also several noisy operations \cite{Genoni_2016}. In particular, in the following we will consider the lossy channel $\mathcal L_\eta$, parametrized by $\eta\in[0,1]$. When applied on the first mode of a two-mode state, it modifies the input covariance matrix $\Gamma$ and displacement vector $\bxi$ as follows:
\begin{equation}\label{def: lossy channel}
\Gamma \stackrel{\mathcal L_\eta^{(A)}}{\longrightarrow} K_\eta \Gamma K_\eta^\intercal + N_\eta N_\eta^\intercal, \qquad \bxi  \stackrel{\mathcal L_\eta^{(A)}}{\longrightarrow} K_\eta \bxi,
\end{equation}
where $K_\eta$ and $N_\eta$ are the block-diagonal matrices
\begin{equation}
K_\eta = \left(\begin{array}{c|c}
\sqrt\eta \Id_2 & \\\hline
& \Id_2
\end{array}\right), \quad
N_\eta = \left(\begin{array}{c|c}
\sqrt{1 \minus \eta} \Id_2 & \\\hline
& \hspace{0.2cm}
\end{array}\right),
\end{equation}
with empty blocks being composed only by zeros.

Due to their simplicity and practical relevance, Gaussian states form the most studied class of density matrices in continuous variable systems. In particular, many quantum information quantifiers can be written in closed form when evaluated on Gaussian states. In the following we will focus on the QFI, which thanks to \Eq{def: QFI as Bures expansion} has been explicitly evaluated for single-mode \cite{Pinel_2013}, two-mode \cite{Safranek_2015}, and for generic multi-mode Gaussian states \cite{Banchi_2015}. 
These formulas, or a perturbative approach based on \Eq{def: QFI as Bures expansion}, have been recently employed in order to assess the performance of Gaussian states in various estimation tasks \cite{Safranek_16_OptGauss,Friis_15}.    
We report here the expression for the QFI of a two-mode probe $\rho_{AB}$ \cite{Safranek_2015}:
\begin{align}\label{eq: two-mode GaussianQFI}
H_\epsilon [\rho_{AB}]& = \frac{1}{2\left(|\tilde M| - 1\right)}\left\{|\tilde M| \Tr{(\tilde M^{-1}\dot{\tilde M})^2}  \right. \\
& + \left. \sqrt{|\Id + \tilde M^2|} \,\Tr{\left((\Id + \tilde M^2)^{-1}\dot{\tilde M}\right)^2}\right\}\notag\\
& + \frac{4}{2(|\tilde M|-1)}(\tilde \nu_1^2 - \tilde \nu_2^2)\left(-\frac{\dot{\tilde \nu}_1^2}{\tilde \nu_1^4-1}+\frac{\dot{\tilde \nu}_2^2}{\tilde \nu_2^4-1}\right) \notag\\
& + 2 \dot{\tilde \bxi}^\intercal \tilde \Gamma^{-1}\dot{\tilde \bxi},\notag
\end{align}
where the symbol $|\cdot|$ represents the determinant, the dot corresponds to the derivative with respect to $\epsilon$, and we defined $M = i \Omega \Gamma$. The tilde appearing on top of all quantities reminds us that they have to be evaluated on the encoded state $\Phi_\epsilon[\rho_{AB}]$.
This formula has a first contribution which depends on the whole covariance matrix, a second one which explicitly takes into account the variation of the symplectic eigenvalues, and a third one which accounts for changes in the displacement vector of the encoded state.
In particular, we point out that the second contribution can lead to irregular behaviors when the encoded state $\rho_\epsilon$ has at least one eigenvalue equal to $1$ \cite{Safranek_2015,Safranek_2016}. 
In what follows we can safely ignore this problem because we will consider either (i) unitary encodings, which do not change the symplectic eigenvalues, or (ii) mixed encoded states with no symplectic eigenvalue equal to $1$.

In the last part of this section, we explicitly consider a local encoding $\Phi_\epsilon = \Phi_\epsilon^{(A)}\otimes\Id_B$ acting nontrivially only on subsystem $A$, and simplify \Eq{eq: two-mode GaussianQFI} to obtain the QFI of a single-mode Gaussian state $\rho_A$. We will get an expression which is equivalent, but not identical, to the one found in Ref. \cite{Pinel_2013}. This alternative expression will be of great help in the following analysis. Let us start by considering a single-mode channel $\Phi_\epsilon^{(A)}$, and its two-mode extension $\Phi_\epsilon^{(A)}\otimes\Id_B$. The latter can be considered a proper two-mode encoding CPT map, so we can apply \Eq{eq: two-mode GaussianQFI} to a separable probe state of the form $\rho_A\otimes\rho_{th}(\nu_B)$, where $\rho_{th}(\nu_B)$ is a thermal state, characterized by $T = \Id_2$ and $D = \nu_B \Id_2$ in \Eq{def: Williamson decomposition}. Thanks to the factorized structure of probe and channel, no change in precision can possibly arise from choosing a different value for $\nu_B$. However, the obtained expression for $H_\epsilon[\rho_A\otimes\rho_{th}(\nu_B)]$ still depends on $\nu_B$ in a nontrivial way. 
This dependence is a consequence of the mathematical structure of \Eq{eq: two-mode GaussianQFI}, but we know that physically the parameter $\nu_B$ cannot play a role in the QFI. Therefore, we can use the trick of considering a factorized probe $\rho_A\otimes\rho_{th}(\nu_B)$ in order to easily deduce a relation between $\tilde M_A$ and its derivative, where $\tilde M_A$ is evaluated on the encoded single-mode state $\Phi_\epsilon^{(A)}(\rho_A)$.  
After a straightforward manipulation, we can see that the studied QFI is independent from $\nu_B$ if and only if the following relation holds:
\begin{align}\label{eq: singMod QFI relation}
\frac{1}{|\tilde M_A|}&\left(\frac{\text{d}|\tilde M_A|}{\text{d}\epsilon}\right)^2 = |\tilde M_A| \, \Tr{\left(\tilde M_A^{-1}\dot{\tilde M}_A\right)^2}\notag \\
& -(1 - |\tilde M_A|)^2 \Tr{[(\Id + \tilde M_A^2)^{-1}\dot{\tilde M}_A]^2}.
\end{align}
This can now be used to substitute either $\Tr{(\tilde M_A^{-1}\dot{\tilde M}_A)^2}$ or $\left(\frac{\text{d}|\tilde M_A|}{\text{d}\epsilon}\right)^2$ in the remainder of \Eq{eq: two-mode GaussianQFI}, in order to obtain an expression for the QFI of a single-mode probe, labeled by $H_\epsilon^{(1)}$. With the former choice, this becomes
\begin{align}\label{eq: Formal singMod QFI}
H_\epsilon^{(1)} (\rho_A) & \eq \frac{|\tilde M_A| \minus 1}{2} \, \Tr{[(\Id \plus \tilde M_A^2)^{-1}\dot{\tilde M}_A]^2} \notag\\
& \plus \frac{1}{2(|\tilde M_A|^2 \minus 1)}\left(\frac{\text{d}|\tilde M_A|}{\text{d}\epsilon}\right)^2 \plus 2\dot{\tilde \bxi}^\intercal_A \tilde \Gamma_A^{-1} \dot{\tilde \bxi}_A,
\end{align}
where $\tilde \Gamma_A$ and $\tilde \bxi_A$ are respectively the covariance matrix and the displacement vector associated with $\Phi_\epsilon^{(A)}(\rho_A)$.
With respect to the expression of Ref. \cite{Pinel_2013}, this is advantageous in all those situations where $\Tr{[(\Id + \tilde M_A^2)^{-1}\dot{\tilde M}_A]^2}$ is easier to compute than $\Tr{(\tilde M_A^{-1}\dot{\tilde M}_A)^2}$, as in Sec. \ref{sec: noisy} below.

\section{Versatile Gaussian Probes for Squeezing Estimation}\label{sec: model}

\begin{figure}
	\centering
	\includegraphics[scale=0.28]{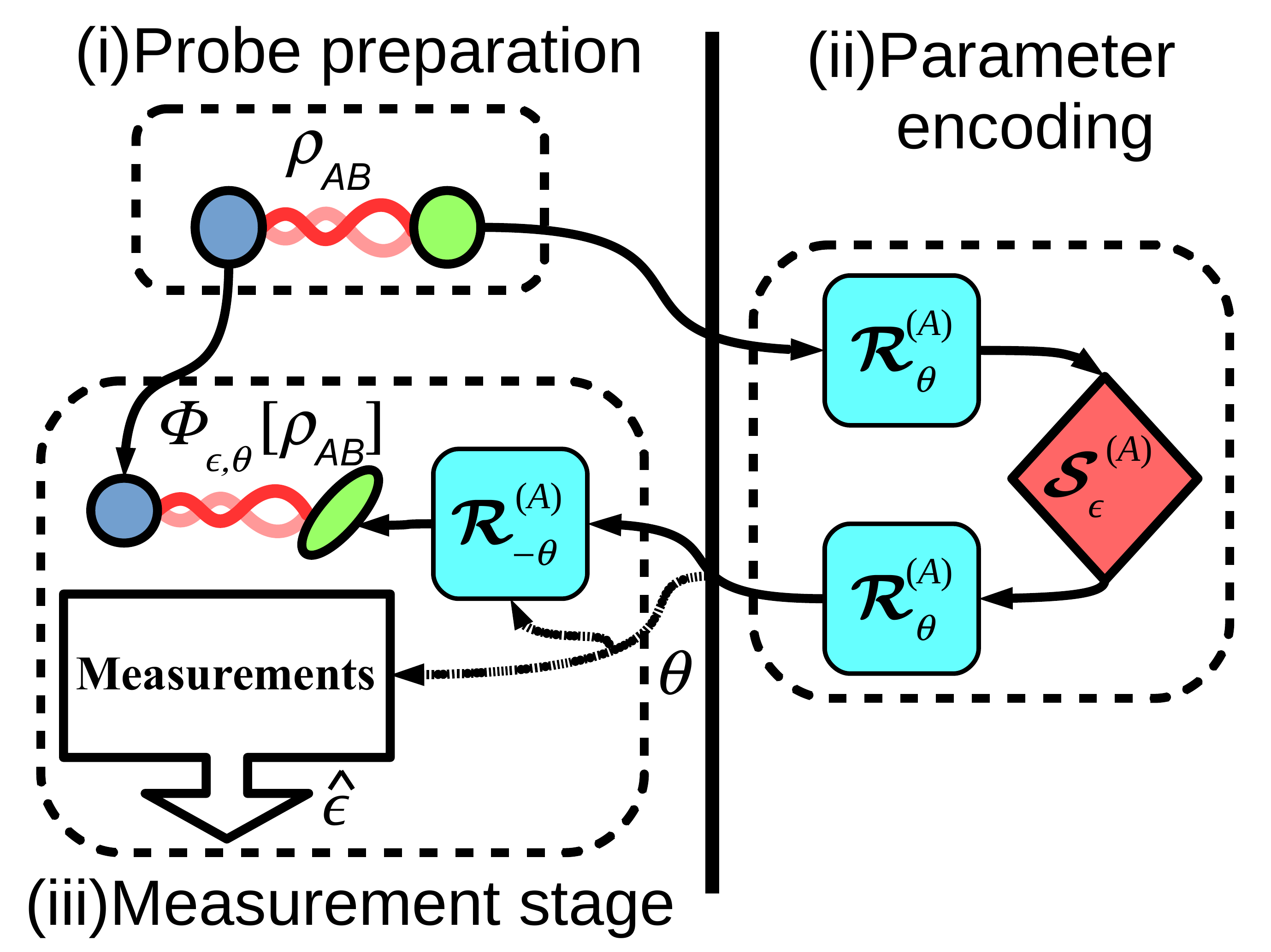}
	\caption{(Color online) Sketch of a physical scenario in which a squeezing operation along an initially unknown direction is applied on the probing system. Stage (i): versatile single-mode or two-mode probes are prepared, without knowing the angle $\theta$ characterizing the estimation process in which they will be used. Stage (ii): the probing systems are sent to the squeezing device, and in each one-way trip they acquire a certain phase $\theta$. Stage (iii): upon recollection of the probes, the angle $\theta$ becomes known, and can be used to devise the best measurement allowed by quantum mechanics, from which an estimator $\hat{\epsilon}$ is recovered. Note that the knowledge of $\theta$ can also be used to apply the unitary correction $\mathcal R_{-\theta}^{(A)}$ on the final state of the probes. Although this does not change the QFI, it allows us to write the total encoding unitary operation in the same form that appears in \Eq{eq: squeezer rotation}.
		\label{fig: sketch}}
\end{figure}

As mentioned in the introduction, we want to study a problem of black-box metrology in which the parameter to estimate is the strength $\epsilon$ of the squeezing applied on the probing state, and an additional uncertainty affects the direction $\theta$ in which the squeezing operation is applied. In absence of noise, the overall encoding CPT map acting on mode $A$ can thus be written as
\begin{equation}\label{eq: squeezer rotation}
\Phi_{\epsilon,\theta}^{(A)} \equiv  \mathcal S^{(A)}_{\epsilon}\circ \mathcal R_{\theta}^{(A)},
\end{equation}
where $\circ$ represents the usual composition of maps. In our analysis we are going to assume that the angle $\theta$ is picked randomly from $[0,2\pi]$ and is unknown at the stage in which the probes are being prepared. Yet  we shall assume that the selected value of $\theta$ is revealed after the parameter $\epsilon$ has been  imprinted into the system. Accordingly, while the  presence of $\theta$ cannot be trivially compensated by properly antirotating the input states of the probes, the knowledge of its value can influence the design of the optimal POVM measurement.
For example, as schematically shown in Fig. \ref{fig: sketch}, this scenario can arise in an optical setup if the distance from a squeezing device, which applies $\mathcal S_{\epsilon}$ to the input state, is not known in advance or is fluctuating from one measurement to the other. In its trip to and from the squeezer, the light will be affected by the free evolution $\mathcal R_\theta^{(A)}$, so that each probe actually evolves via $\mathcal R_\theta^{(A)} \circ\mathcal S_\epsilon^{(A)}\circ \mathcal R_\theta^{(A)}$. Although $\theta$ is not known when the probes are being prepared, in this example its value could be inferred by the light travel time, or it could be independently estimated once the setup has been set, and the estimation experiment is about to be performed. However, it is important to keep in mind that this information can only be used to optimize the readout measurement, and not the states of the probes used in the following experimental runs, because we are assuming that the probing systems have been selected and prepared beforehand.
Upon receiving the encoded state, it is e.g. possible  to add a unitary correction $\mathcal R_{-\theta}^{(A)}$ to $\mathcal R_\theta^{(A)} \circ\mathcal S_\epsilon^{(A)}\circ \mathcal R_\theta^{(A)}$, changing the effective evolution of a single-mode probe $\rho_A$ to match the form of \Eq{eq: squeezer rotation} (further compensations being possible if required by the measurement optimization stage). 

Under the above conditions the quantum Cram\'{e}r-Rao bound (\ref{eq: quantum CRB}) predicts that the ultimate estimation accuracy achievable with a probing state $\rho$ exhibits
an explicit functional dependence upon $\theta$, 
\begin{eqnarray} \label{NEWEQ}
\delta {\hat{\epsilon}}(\theta) \geq \frac{1}{\sqrt{M {H}^{(\theta)}_\epsilon (\rho) }},
\end{eqnarray} 
with $H^{(\theta)}_\epsilon (\rho)$ being the QFI for $\epsilon$ evaluated on $\Phi_{\epsilon, \theta}[\rho]$. It is hence very possible that in the estimation of $\epsilon$ 
an input density matrix $\rho$  will provide different performances   
depending on the  value of $\theta$.  In this context the versatility of an input state can be gauged by  looking at the average QFI (AvQFI) it is capable of 
 granting, i.e., the quantity   
\begin{equation}\label{eq: AvQFI}
\overline{H}_\epsilon (\rho) \equiv \int_0^{2\pi}
\frac{\text{d}\theta}{2\pi} \, 
H^{(\theta)}_\epsilon (\rho).
\end{equation}
From Eq.~(\ref{NEWEQ})  and from the convexity of the  function $1/\sqrt{x}$ it follows that $\overline{H}_\epsilon (\rho)$ sets a lower bound on the 
average value of the attainable RMSEs, i.e.,
\begin{equation}\label{eq: classical CRB}
\overline{\delta {\hat{\epsilon}}}   \equiv \int_0^{2\pi}\frac{\text{d}\theta}{2\pi} \, \delta {\hat{\epsilon}}(\theta)
  \geq 
\frac{1}{\sqrt{M \overline{H}_\epsilon (\rho) }}.
\end{equation}
As explicitly discussed in  Appendix \ref{app: AvQFI fluctuating theta}, the AvQFI quantity  can be used also to bound the accuracy achievable in an alternative setting where,
at variance with the case represented in Fig.~\ref{fig: sketch}, the phase $\theta$ fluctuates over different probing stages.

Although the encoding operation given in \Eq{eq: squeezer rotation} acts only on a single mode of the field, we can nonetheless consider the possibility of using a two-mode probe. In this case, one of its modes (say, $A$) is sent to the squeezing station, while its second mode ($B$) is kept unaltered in the laboratory as reference. As often happens, the presence of initial correlations between the two subsystems could potentially help in the estimation of $\epsilon$, even if mode $B$ is not directly affected by the evolution. In this more general case, the total encoding map can be written as
\begin{align}\label{def: noiseless evolution}
\Phi_{\epsilon,\theta}(\rho_{AB})  &= \Phi_{\epsilon,\theta}^{(A)}\otimes\Id^{(B)}[\rho_{AB}].
\end{align}
In the remainder of this paper we will label by $\overline H_\epsilon$ the AvQFI associated with a two-mode probe. In all those situations where we are explicitly using a single-mode probing system, we will emphasize this choice by using the symbol $\overline H_\epsilon^{(1)}$ for the AvQFI. 

It is now worthwhile to briefly discuss the properties of the AvQFI.
In particular, it is convex in the input probe because it inherits this property from the QFI. The AvQFI is also invariant under phase rotations on $A$ and generic unitary evolutions $\mathcal U^{(B)}$ applied on mode $B$, i.e.,
\begin{equation}\label{eq: AvQFI symmetry}
\overline{H}_\epsilon\left[\mathcal R_\phi^{(A)}\otimes \mathcal U^{(B)} [\rho_{AB}] \right] = \overline{H}_\epsilon \left[\rho_{AB}\right].
\end{equation}
This can be seen in two steps. At first, $\mathcal U^{(B)}$ commutes with the encoding channel $\Phi_{\epsilon,\theta}$, and it can be absorbed in the measurement process, thus leaving the QFI unaltered. Then, notice that 
\begin{equation}
\Phi_{\epsilon,\theta} \circ\mathcal R_\phi^{(A)} [\rho_{AB}] = \mathcal R_\phi^{(A)} \circ\Phi_{\epsilon,\theta+\phi}\left[\rho_{AB}\right] .
\end{equation}
Once again, the external $\mathcal R_\phi^{(A)}$ can be included in the readout process, while the shift in $\theta$ vanishes because of the average appearing in \Eq{eq: AvQFI}.
Crucially, we can exploit the symmetry of \Eq{eq: AvQFI symmetry} in order to simplify the structure of the set of probes $\rho_{AB}$ that we need to explicitly consider in our search for the optimal average performance. As detailed in appendix \ref{AppStd}, it is enough to consider states in the standard form $\rho^{\text{(std)}}$ characterized by:
\begin{equation}\label{def: std parametrization}
\Gamma^{\text{(std)}} \eq \left(\begin{array}{cc|cc}
a_x & \!a_{xp}\! & c & 0 \\
\!a_{xp}\! & a_p & 0 & d\\\hline
c & 0 & b & 0 \\
0 & d & 0 & b
\end{array}\right), \quad \bxi^{\text{(std)}} \eq (\xi_x,\xi_p,0,0)^\intercal.
\end{equation}

Equation~\eqref{def: std parametrization} is a good parametrization for two-mode input states, but whenever we deal with single-mode probes it is convenient to use a different approach. The covariance matrix and displacement vector of a generic single-mode Gaussian state $\rho_A$ can be decomposed as
\begin{equation}\label{eq: single-mode Gauss state}
\Gamma_A = \nu_A R_\phi S_{2\alpha} R_\phi^\intercal,  \quad  \bxi_A = |\xi| \left(\begin{array}{c}
\cos\psi\\
\sin\psi
\end{array}\right),
\end{equation} 
where $\nu_A$ characterizes its thermal excitations, $\alpha$ and $\phi$ quantify respectively the amount and the direction of squeezing, while $\psi$ fixes the displacement direction. 
After the application of the phase rotation $\mathcal R_{\theta}^{(A)}$, the parameters $\phi^\prime$ and $\psi^\prime$ of the evolved state $\mathcal R_{\theta}^{(A)} [\rho_A]$ are respectively $\phi^\prime = \phi + \theta$ and $\psi^\prime = \psi - \theta$. As the average over $\theta$ in the AvQFI can be equivalently performed over $\theta + \phi$, only the sum $\phi + \psi$ can influence the average estimation precision obtained with the probe $\rho_A$. For this reason, without loss of generality in the following we can set $\phi = 0$ in \Eq{eq: single-mode Gauss state} when parametrizing single-mode probes.

Although here we explicitly discussed the noiseless unitary encoding given in \Eq{def: noiseless evolution}, we point out that the same reasoning that led us to Eqs. \eqref{def: std parametrization} and \eqref{eq: single-mode Gauss state} can be applied also in Sec.\ref{sec: noisy}, where we consider a noisy evolution. This is because we only deal with photon losses, whose CPT map $\mathcal L_\eta$, defined in \Eq{def: lossy channel},
commutes with phase rotations.

\subsection{Results for noiseless evolution}\label{sec: noiseless}
In order to find the average performance of a two-mode Gaussian probe state for the estimation of the squeezing parameter $\epsilon$, characterizing the noiseless evolution $\Phi_{\epsilon,\theta}$ defined in \Eq{def: noiseless evolution}, we first need to evaluate its $\theta$-dependent QFI through \Eq{eq: two-mode GaussianQFI}. We stress that due to the unitarity of the evolution, we can ignore the contribution coming from the derivatives of the symplectic eigenvalues.
The remaining terms can be explicitly evaluated for input states of the form of  \Eq{def: std parametrization} by writing the matrix $\tilde M$ in terms of the input covariance matrix $\Gamma$ as follows:
\begin{equation}\label{eq: noiseless term 1}
\Tr{\left(\tilde M^{-1} \dot{\tilde M} \right)^2}=2 \,\Tr{ \Gamma^{-1}V_\theta  \Gamma V_\theta +V_\theta^2},
\end{equation}
\begin{align}\label{eq: noiseless term 2}
\Tr{\left((1+\tilde M^2)^{-1}\dot{\tilde M} \right)^2} = & -\Tr{ (O V_\theta \Gamma + O \Gamma V_\theta )^2 },
\end{align}
where $V_\theta = R_\theta^\intercal (S_\epsilon^{-1} \dot S_\epsilon)R_\theta$ and $O = (\Id-\Omega\Gamma \Omega \Gamma)^{-1}\Omega$.
An analytical, quite involved, expression for the AvQFI can be found in Appendix~\ref{app: Noiseless two-mode AvQFI}.

By setting $c,d = 0$ we can obtain a simpler expression that does not depend on $b$, which can be interpreted as the average performance of a single-mode probe. Overall, its average QFI can be written as:
\begin{equation}\label{eq: SM AvQFI}
\overline {H}^{(1)} [\rho_A] = \frac{\Tr{\Gamma_A}^2 + 4\det\Gamma_A}{2 (1 + \det \Gamma_A)} + \frac{|\xi|^2\Tr{\Gamma_A}}{\det \Gamma_A},
\end{equation}
where $\Gamma_A$ is the input covariance matrix and ${|\xi|=\sqrt{\xi_x^2 + \xi_p^2}}$.
We begin by commenting the single-mode result, and then we move to study the effects of input correlations.

\subsubsection{Single-mode probes}
Remarkably, \Eq{eq: SM AvQFI} is independent from the phase $\psi$ appearing in \Eq{eq: single-mode Gauss state}, and the displacement only appears through its absolute value $|\xi|$.  This is a peculiar characteristic of the considered noiseless evolution, which disappears when we take losses into account in Sec. \ref{sec: noisy}. We note that the single-mode AvQFI of \Eq{eq: SM AvQFI} can alternatively be obtained by averaging the single-mode QFI for fixed squeezing direction found in Ref. \cite{Safranek_2015}. This can be easily shown by writing their QFI in our notation, for input states parametrized as in \Eq{eq: single-mode Gauss state} with $\phi = 0$:
\begin{align}\label{eq: theta SM QFI}
& H^{(1)}_{\theta} = \frac{2 |\xi|^2}{\nu_A}\left(\cosh(2\alpha) +\cos[4\theta -2\psi]\sinh(2\alpha)\right) \notag \\
& \plus \frac{4 \nu_A^2}{\nu_A^2 \plus 1}\left( \! \cosh^4\alpha \plus \sinh^4\alpha \minus \frac{1}{2}\cos[4\theta]\sinh^2(2\alpha) \! \right).
\end{align}
\Eq{eq: SM AvQFI} can then be retrieved from this expression by averaging over $\theta$.

\begin{figure}
	\centering
	\includegraphics[scale=0.6]{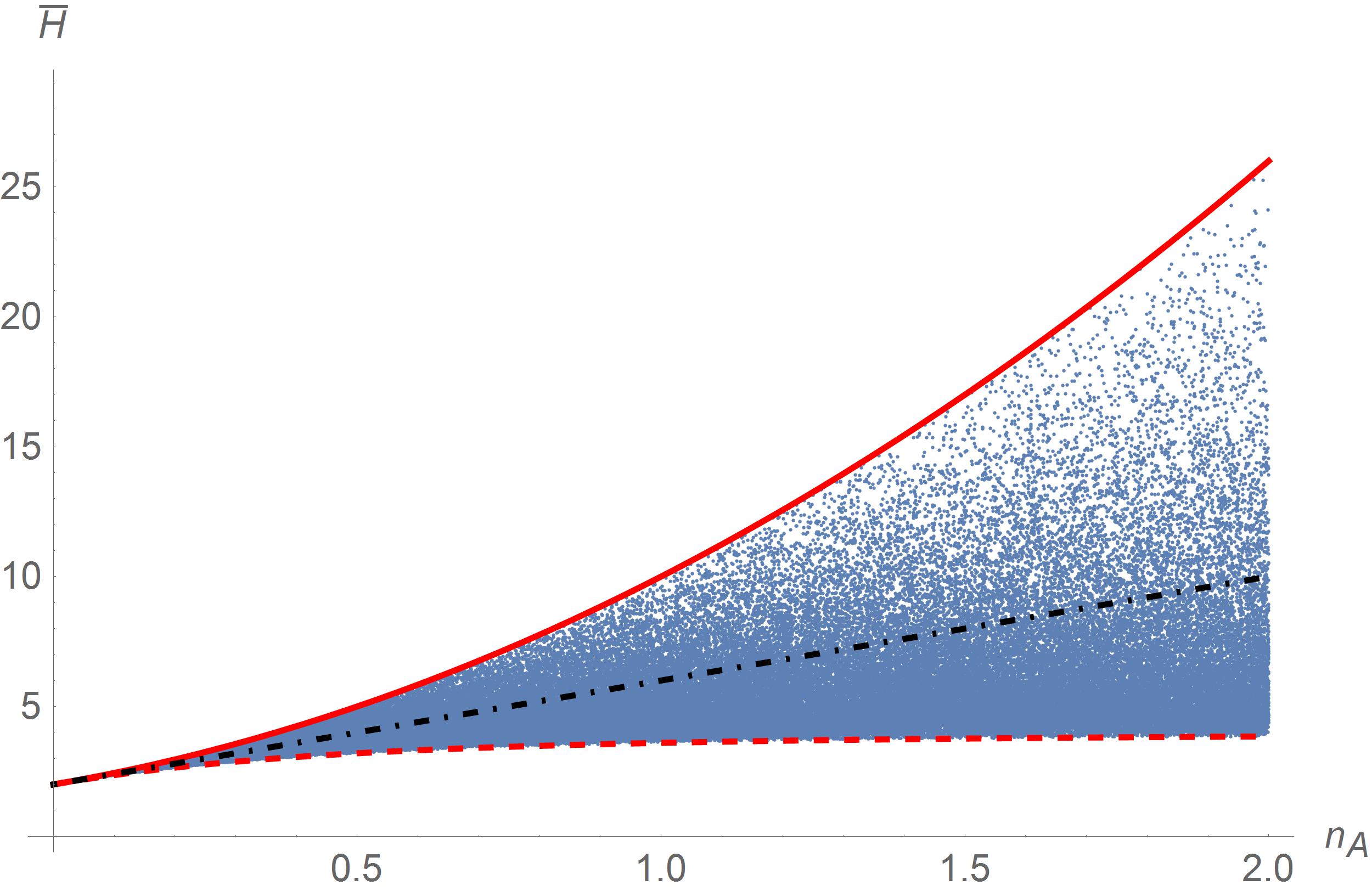}
	\caption{(Color online) Noiseless AvQFI for single-mode Gaussian probes with respect to the photon number $n_A$. Blue dots: $10^5$ single-mode Gaussian state uniformly sampled with the method of Appendix \ref{app: sampling appendix}; red top solid line: pure undisplaced squeezed states; red bottom dashed line: undisplaced thermal states; black dot-dashed line: coherent states. \label{fig: noiseless AvQFI}}
\end{figure}

Since we are dealing with states defined in an infinite-dimensional Hilbert space, we take the physically meaningful approach of looking for the optimal probe under the condition of fixed input energy, i.e., fixed average photon number 
\begin{equation}\label{def: formula n_A}
n_A = \frac{\nu_A \cosh(2\alpha) -1 + |\xi|^2}{2}.
\end{equation}
This is because an arbitrary amount of energy can lead to an unbounded estimation precision.
A plot of $\overline {H}^{(1)} (\rho_A)$ against $n_A$ for randomly generated single-mode input Gaussian states can be found in Fig.~\ref{fig: noiseless AvQFI}. We can see that the optimal probe is given by a pure undisplaced squeezed state ($|\xi| = 0$, $\nu_A = 1$), while the worst performance is obtained when an undisplaced thermal state is used ($|\xi| = 0$, $\alpha = 0$). The corresponding AvQFIs are respectively given by:
\begin{align}
\overline H^{(1)} \left[\rho_A^{(\text{sq})}\right]& = 4 n_A^2 + 4 n_A + 2,\\
\overline H^{(1)} \left[\rho_A^{(\text{th})}\right]& = 4 \frac{(2 n_A + 1)^2}{1+(2 n_A + 1)^2},
\end{align}
while coherent states ($\alpha=0$, $\nu_A = 1$) have an intermediate scaling
\begin{equation}
\overline H^{(1)} \left[\rho_A^{(\text{coh})}\right] = 2(1+2 n_A).
\end{equation}
Formal proofs of these bounds can be found in Appendix~\ref{app: noiseless bounds}.

\begin{figure}
	\centering
	\includegraphics[scale = 0.6]{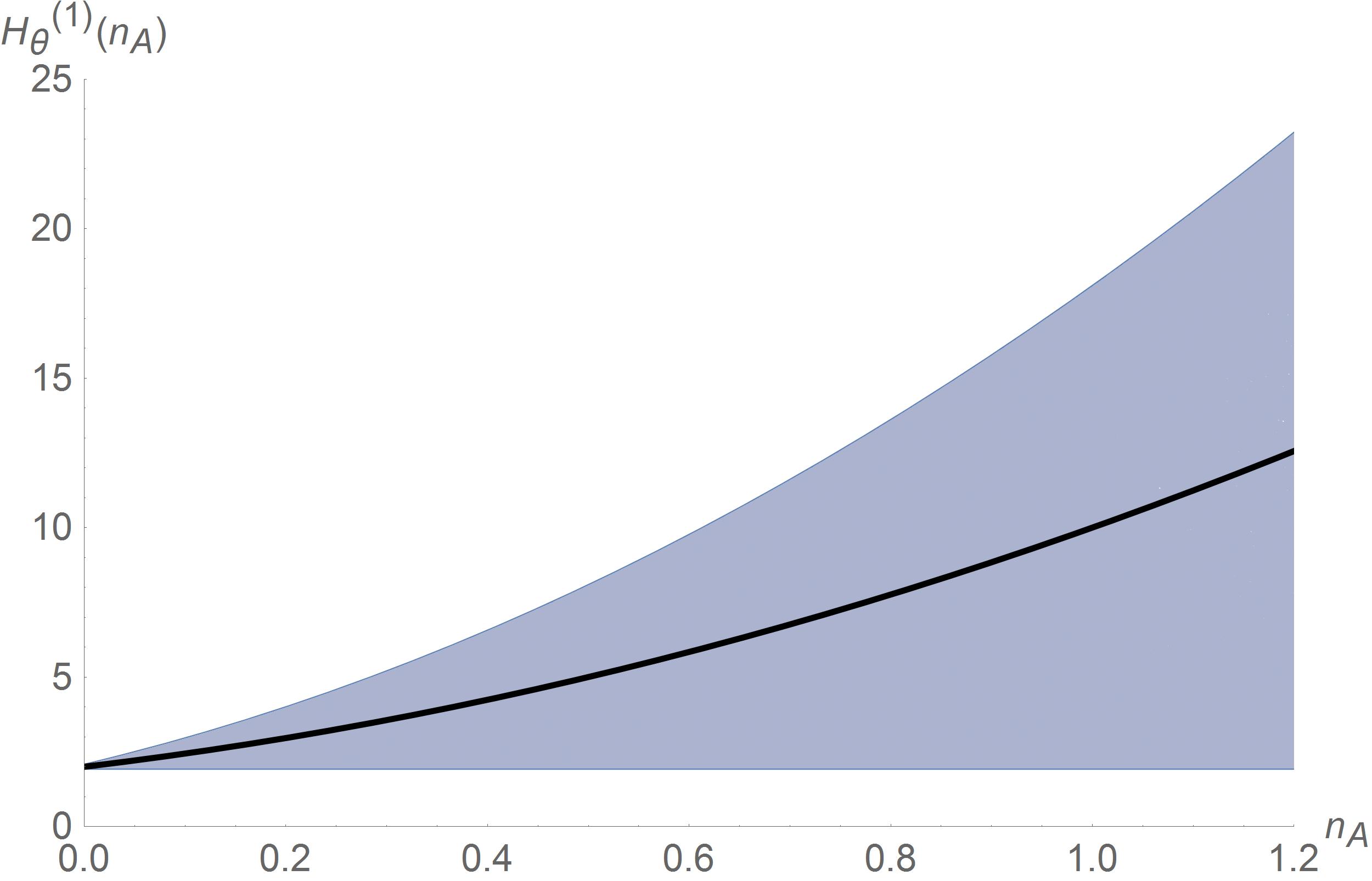}
	\caption{(Color online) Range of QFI values that could be achieved for any fixed $n_A$ by varying $\theta$, for pure and undisplaced squeezed single-mode probes $\rho_A^{(\text{sq})}$. The corresponding AvQFI is plotted as a black line for comparison.} \label{fig: variance}
\end{figure}

For single-mode input probes, we can also study the variance over $\theta$ of the QFI written in \Eq{eq: theta SM QFI}. This is found to be
\begin{equation}
\text{VAR}(H_\theta) = \frac{V_1^2 + V_2^2 - 2 V_1 V_2 \cos(2\psi)}{2},
\end{equation}
where 
\begin{equation}\label{eq: variance}
V_1 = \frac{2 \nu_A^2}{\nu_A^2 + 1} \sinh^2(2 \alpha), \quad
V_2 = \frac{2 |\xi|^2}{\nu_A} \sinh(2\alpha).
\end{equation}
As can be expected, when $\xi = 0$ the variance does not depend on $\psi$. Moreover, it is identically zero when $\alpha = 0$, or when the probe displacement is opportunely chosen so as to have $V_1 = V_2$, with $\psi$ an integer multiple of $\pi$. 
Therefore, the QFI of $\rho_A^{(\text{th})}$ and $\rho_A^{(\text{coh})}$ does not depend on $\theta$, as they are characterized by ${\alpha = 0}$. On the contrary, the variance in \Eq{eq: variance} for pure undisplaced squeezed states increases with $\alpha$, and thus with $n_A$. This implies that the single-mode probes associated with the optimal average performance also yield strong fluctuations in precision with respect to $\theta$. Although the QFI associated with these probes is non-zero for any $\theta$, an unlucky choice of $\theta$ can bring the estimation precision to its absolute lower bound (QFI = 2) (see Fig. \ref{fig: variance}).

\subsubsection{Two-mode probes}
We can now check what changes if we use two-mode Gaussian probes, in which only the first mode goes through the squeezing device while the second one is kept as a reference. In this framework, we can compare the AvQFI of different input states either by fixing the number of photons $n_A$ that go through the squeezing device, or by fixing the total number of photons $N$. Let us start with the former comparison. A simple observation is that in this case any two-mode Gaussian state does not perform worse than its single-mode reduction $\rho_A = \TrS{\rho_{AB}}{B}$, because the QFI is monotonically decreasing under partial trace. This trivially implies:
\begin{equation}
\overline H[\rho_{AB}] \geq \overline H^{(1)}[\rho_A] \geq \min_{\rho_A} \overline H^{(1)}[\rho_A].
\end{equation}
\emph{A priori}, it might be possible to improve the average estimation precision by exploiting correlations with an ancillary mode, for example by using input entangled states. An obvious candidate in looking for this sort of advantage would be a correlated two-mode extension of the optimal single-mode probe. However, this state does not exists because $\rho_A^{(\text{sq})}$ is pure and cannot be correlated with any other system. 
This suggests the presence of a trade-off between pure local squeezing and two-mode correlations, consistently with the results known in the finite-dimensional case \cite{Farace_2016}.
As a paradigmatic example of correlated probes, we study the class of two-mode squeezed vacuum states $\rho_{AB}^{(\text{sq})}(r)$, usually considered as the continuous variables counterpart of maximally entangled states. In the notation of \Eq{def: std parametrization}, their standard parameters are $\xi_x = \xi_p = a_{xp} =  0$ and
\begin{equation}\label{eq: twim-beam parameters}
a_x = a_p = b = \cosh(2r), \qquad c = - d = \sinh(2r).
\end{equation}
In this case, the general formula for $\overline H(\rho_{AB})$ given in  Appendix~\ref{app: Noiseless two-mode AvQFI} greatly simplifies to
\begin{equation}\label{eq: twim-beam QFI}
\overline H[\rho_{AB}^{(\text{sq})}] = 4 \sinh(r)^4 + 4 \sinh(r)^2 + 2.
\end{equation}
Interestingly, for this class of states the number of photon in mode $A$ is equal to $\sinh(r)^2$ and we see that 
\begin{equation}\label{eq: noiseless optimum}
\overline H[\rho_{AB}^{(\text{sq})}] = \overline H^{(1)}[\rho_{A}^{(\text{sq})}] = \max_{\rho_A} \overline H^{(1)}[\rho_A].
\end{equation}
Therefore, $\rho_{AB}^{(\text{sq})}$ yields the same average performances of a pure single-mode squeezed probe. 
This differs from what has been recently found in a finite-dimensional setting \cite{Farace_2016}, even tough exploiting the average skew information \cite{Wigner_SkewInfo_1963,Luo_SkewInfo_2003} as figure of merit, where entanglement was necessary in order to obtain the maximum average precision.
We have strong numerical evidences that all other two-mode Gaussian states with the same $n_A$ yield worse average precisions than $\rho^{(\text{sq})}_{AB}$. 
Another interesting observation is that the QFI associated with $\rho_{AB}^{(\text{sq})}$ is constant over all choices of $\theta$ and equal to its average in \Eq{eq: twim-beam QFI}. Indeed, due to the symmetry of their covariance matrix [see \Eq{eq: twim-beam parameters}], all $\theta$ dependencies in \Eq{eq: noiseless term 1} and \Eq{eq: noiseless term 2} cancel. Therefore, even if single-mode and two-mode squeezed states lead to the same average performances with respect to $n_A$, the latter choice removes fluctuations at the cost of doubling the total number of photons $N$ in the probe. Input correlations are therefore beneficial in all those situations where the guarantee of obtaining a certain predictable performance is preferable to the risk of dealing with fluctuations in estimation precision.

Finally, we still have to discuss what happens if we compare the AvQFIs of two-mode and single-mode probes with the same total number of photons $N$. From \Eq{eq: noiseless optimum} and the numerical evidences in support of the fact that $\rho_{AB}^{(\text{sq})}$ seems to be the best two-mode probe, it should be clear that with this meter of comparison the presence of a reference beam cannot improve the estimation precision. Indeed, two-mode squeezed states can match the performance of single-mode squeezed states only at the cost of doubling the total photon number.

\subsection{Results for noisy evolution}\label{sec: noisy}
Up to now, we considered the ideal and noiseless evolution $\Phi_{\epsilon,\theta} $ defined in \Eq{def: noiseless evolution}. In this section we move to a more realistic scenario by introducing some noise in the picture. In particular, we consider photon losses occurring during the propagation of the probes to and from the squeezing device. The resulting encoding channel acting on subsystem $A$ can be written as:
\begin{equation}\label{def: noisy channel}
\tilde \Phi_{\epsilon,\eta,\theta}^{(A)} = \mathcal L_\eta^{(A)} \circ \Phi^{(A)}_{\epsilon,\theta} \circ \mathcal L_\eta^{(A)},
\end{equation} 
where the action of the lossy channel has been detailed in \Eq{def: lossy channel}. We consider the loss parameter $\eta$ to be fixed and known. Differently from before, the physical map that encodes the parameter $\epsilon$ on the probing system is not unitary. This fact has two main consequences: in general the QFI will be $\epsilon$-dependent, and the possible changes in the symplectic eigenvalues of the probe contribute to the QFI via the third line of \Eq{eq: two-mode GaussianQFI}. 

We will show that, in a noisy environment, correlated two-mode probes can lead to higher average precisions than the optimal single-mode input states. It turns out that this is always true when we compare AvQFIs of states with the same $n_A$. Interestingly, the same result can hold even if the comparison is performed by fixing the \emph{total} photon number $N$ of the  probe state, if the pair $(N,\eta)$ lies within a certain region.

\subsubsection{Optimal single-mode probes} 
In analogy with the noiseless analysis, we are able to find a closed expression for the single-mode QFI in presence of losses. This is one of those cases where our expression for the single-mode QFI, given in \Eq{eq: Formal singMod QFI}, results in being useful. Indeed, when we apply the encoding channel $\tilde \Phi_{\epsilon,\eta,\theta}^{(A)}$ to a probe with standard covariance matrix as in \Eq{def: std parametrization}, $(\Id + \tilde M_A^2)$ becomes a multiple of the identity. Its inverse is therefore much easier to compute than the inverse of $\tilde{M}_A$, which would be required if the expression given in Ref. \cite{Pinel_2013} were to be used.
The explicit expression for the $\theta$-dependent QFI can be found in Appendix \ref{app: expressions noisy QFI} for any values of $\eta$ and $\epsilon>0$. 
Due to the complexity of the obtained expression, we cannot analytically average it over $\theta$ in $[0,2\pi]$, but we can study it numerically. 

A first interesting feature is that a dependence on $\psi$ is generally retained even after the average over the squeezing direction $\theta$ is performed. This is in contrast with the noiseless case, where only the absolute value of the displacement was relevant [see \Eq{eq: SM AvQFI}]. 
Without loss of generality we can still fix $\phi = 0$ in the state parametrization given by \Eq{eq: single-mode Gauss state}. Then, for any fixed values of $n_A, \epsilon,\eta$, and $\alpha$ we find that the maximum of $\overline H_{\epsilon,\eta}$ is reached when $\psi = \pm \pi/2$. This corresponds to a displacement in the direction of the quadrature with the smallest variance. This fact is formally proven Appendix \ref{app: expressions noisy QFI}.

\begin{figure}
	\centering
	\includegraphics[scale = 0.8]{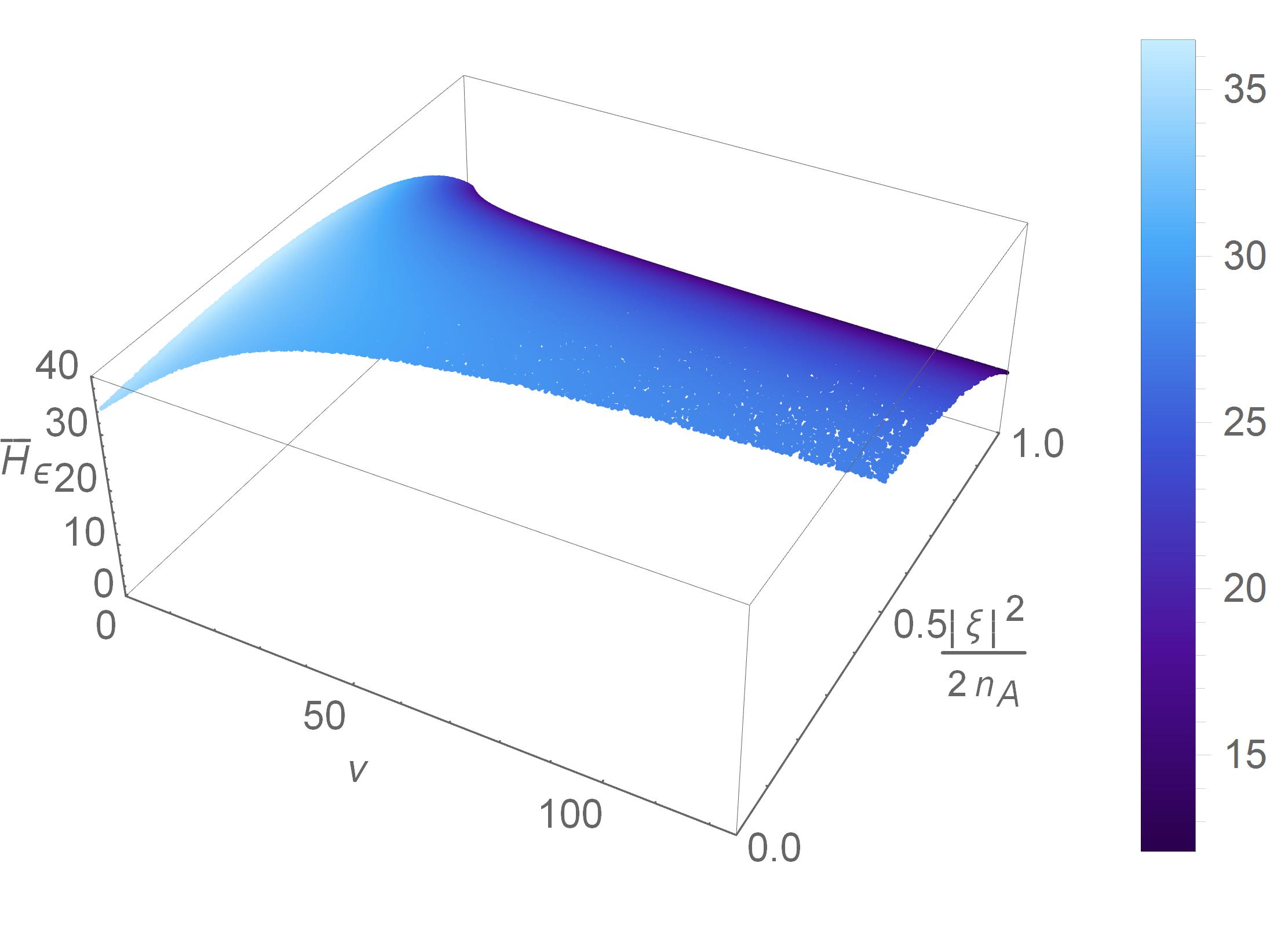}
	\caption{(Color online) AvQFI with respect to $\nu$ and $|\xi|^2/(2 n_A)$ for $\epsilon=1$, $n_A =5$, and $\eta = 0.95$. We uniformly sampled $10^5$ random Gaussian states with $\phi = 0$ and $\psi = \pi/2$, according to the method detailed in Appendix \ref{app: sampling appendix}. \label{fig: AvQFI 3D plot}}
\end{figure}

With this optimal choice for $\psi$, we can parametrize all single-mode probes through the parameters $n_A$, $\nu$ and $|\xi|^2$ [$\alpha$ is then uniquely determined from the energy constraint of \eqref{def: formula n_A}]. We have strong numerical evidences that pure states (i.e., $\nu=1$) seem to reach the maximum AvQFI value among all single-mode probes, for any fixed value of $n_A$ and $\eta$. We point out that convexity of AvQFI does not allow us to prove this conjecture because of the constraint on the average photon number of the probe. In Fig.~\ref{fig: AvQFI 3D plot} we can see a typical plot showing the dependence of AvQFI on $\nu$ and $|\xi|^2/(2 n_A)$, for $\epsilon =1$, $n_A = 5$ and $\eta = 0.95$. 

\begin{figure}
	\centering
	\includegraphics[scale = 0.7]{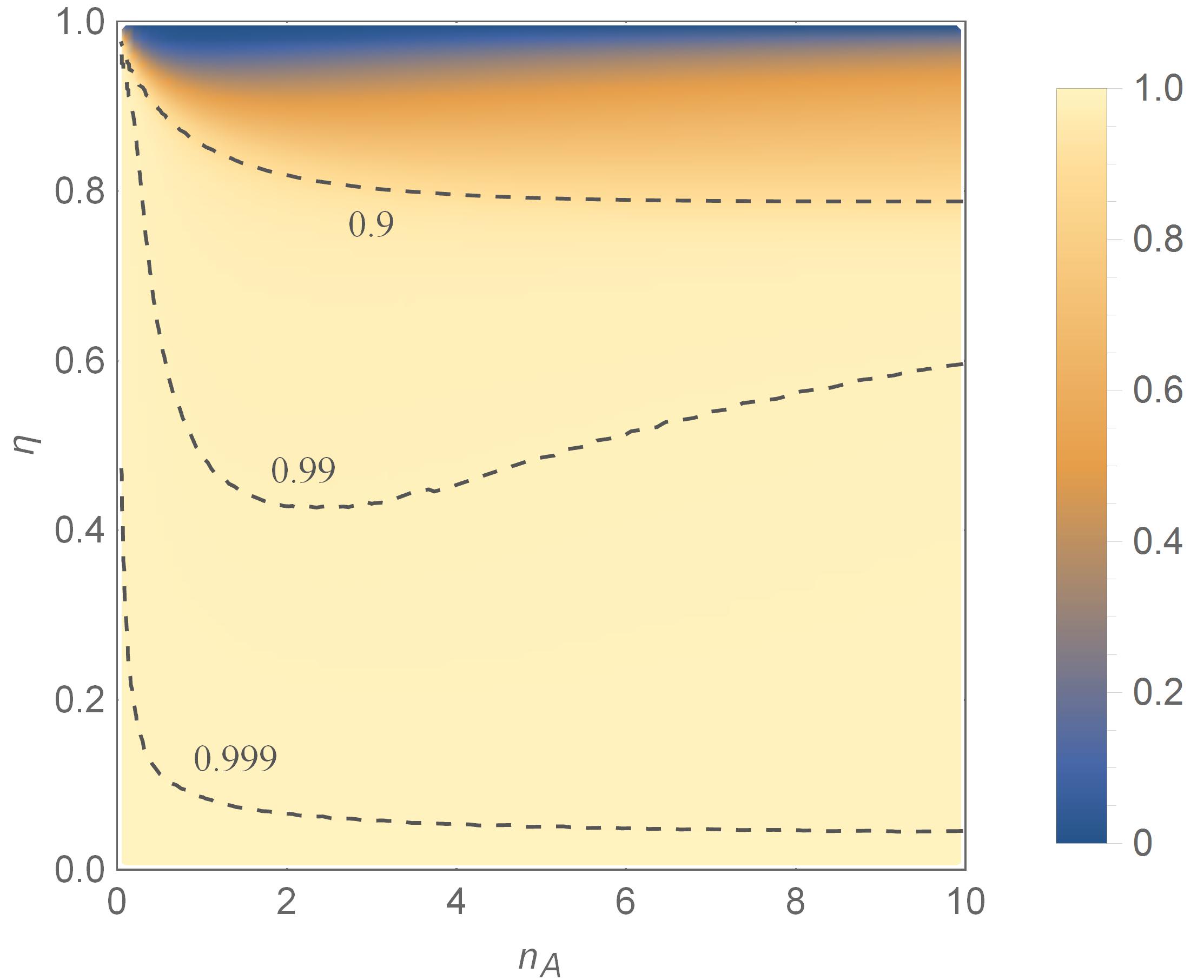}
	\caption{(Color online) Optimal value for the ratio $|\xi|^2/(2 n_A)$, leading to the maximum AvQFI value for a fixed pair $(n_A,\eta)$. The plot is obtained by considering $\epsilon = 1$. \label{fig: optimal single-mode probe}}
\end{figure}

In searching for the optimal single-mode state, therefore, we fix $\nu = 1$ and numerically look for the ratio $|\xi|^2/(2 n_A)$ that yields the largest AvQFI around a given value of $\epsilon$ (remember that the problem is $\epsilon$-dependent in the noisy case). Intuitively, this ratio tells us the percentage of photons that we should use for displacing the state, rather than for squeezing. We plot the result in Fig.~\ref{fig: optimal single-mode probe}, for $10^4$  pairs $(n_A,\eta) \in [0,10]\times[0,1]$, for the specific example of $\epsilon = 1$.  
We already know that for vanishing losses ($\eta=1$) the best strategy is to squeeze the input state as much as possible, and we retrieve this feature from the plot. However, we also see that for high losses the opposite choice leads to better average performances. In particular, for a wide range of values of $\eta$ below a certain $n_A$-dependent threshold, the optimal probe can be considered a coherent state for all practical purposes (although strictly speaking a vanishingly small squeezing component is always required). In the intermediate regime, the optimal probe has a non-zero amount of both squeezing and displacement. If the value of $\epsilon$ is decreased, we obtain a similar behavior, but with a much faster transition between the two extreme regimes where the optimal parameter $|\xi|^2/(2 n_A)$ is $0$ or $1$ (see Appendix \ref{app: epsilon plots} for the plots associated with $\epsilon = 0.5$ or $0.1$).

\subsubsection{Comparison with correlated probes}
\begin{figure}
	\centering
	\subfloat[Comparison for fixed $n_A$. \label{fig: nA comparison}]{
		\includegraphics[scale=0.7]{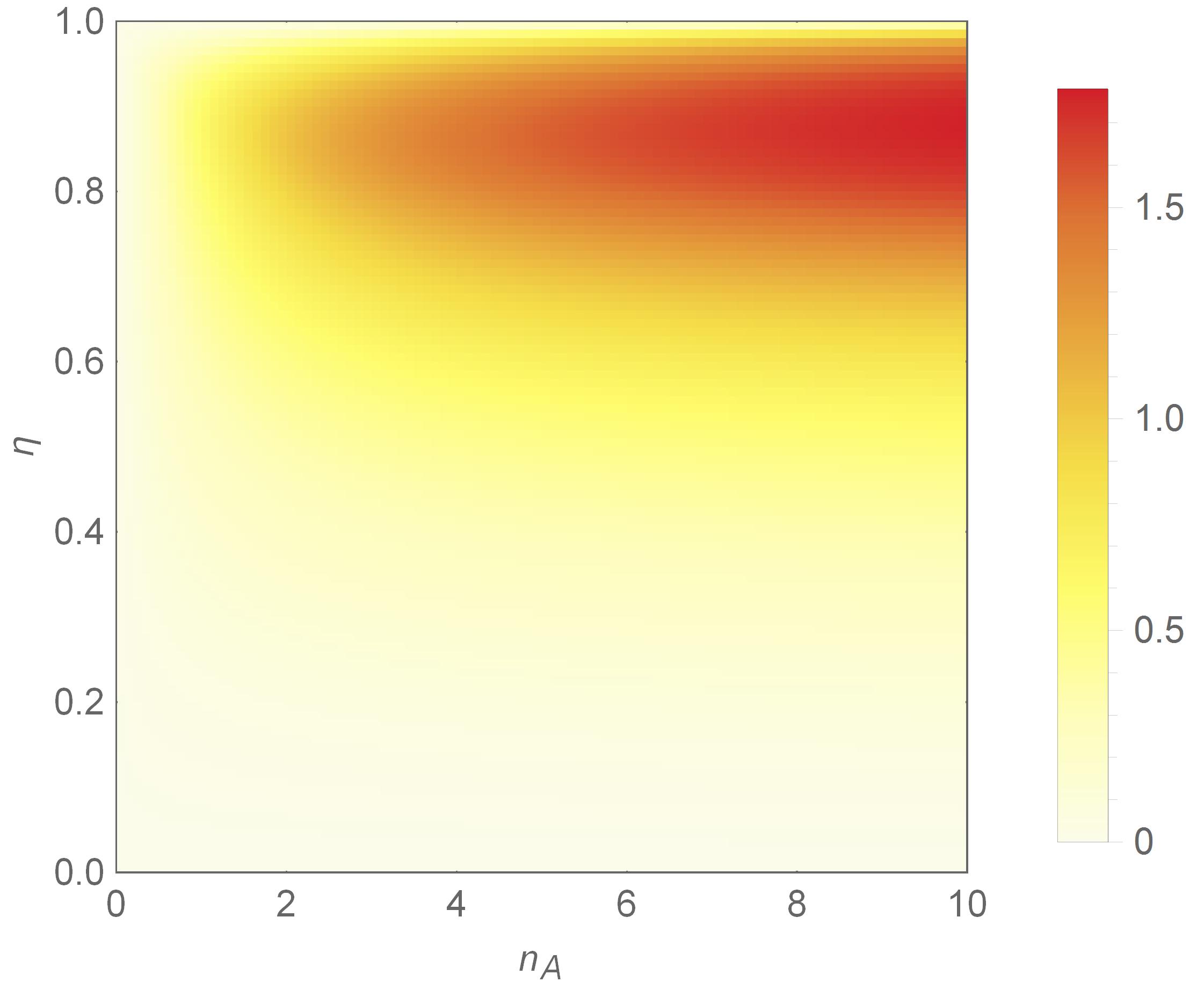}}\\
	\subfloat[Comparison for fixed $N$. \label{fig: N comparison}]{
		\includegraphics[scale=0.7]{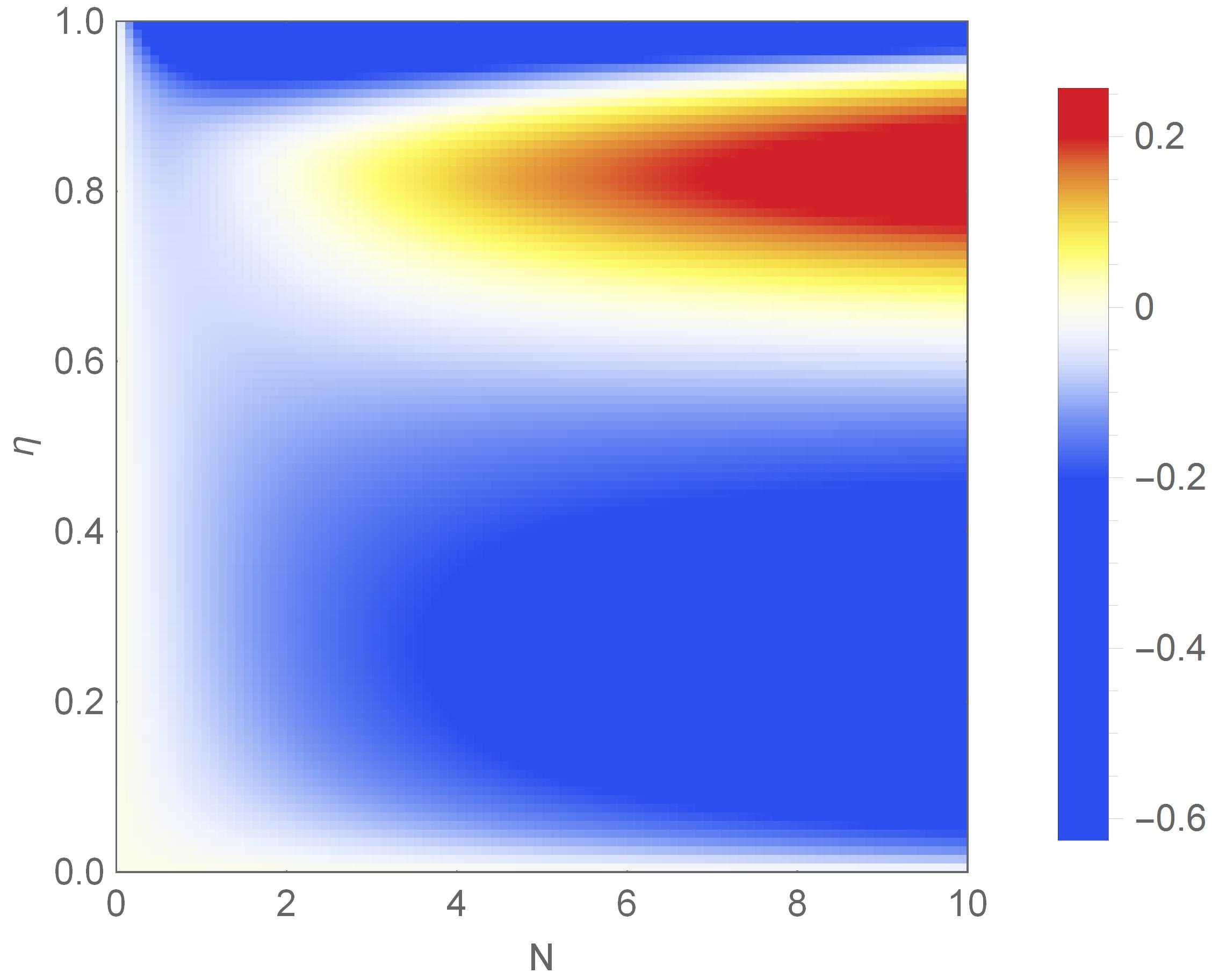}}\\
	\caption{(Color online) Relative increase $\mathcal I$ in precision obtained by using two-mode squeezed vacuum probes rather than the optimal single-mode input state, numerically evaluated for $10^4$ pairs $(n_A, \eta)$ or $(N, \eta)$ when $\epsilon = 1$. \label{fig: two-mode comparison}}
\end{figure}

The AvQFI for the noisy encoding can be calculated for two-mode squeezed vacuum probes. This is a specific but paradigmatic choice; indeed from the results of the noiseless case we can reasonably expect that two-mode squeezed vacuum states remain optimal. The results obtained for the particular choice $\epsilon = 1$ and different values of $n_A$ (or total photon number $N$) and $\eta$ can then be compared with the largest AvQFI obtainable by using single-mode probes with the same photon number. In Fig.~\ref{fig: two-mode comparison} we plot the relative increase in precision that can be obtained by using this correlated input state, namely 
\begin{equation}
\mathcal I = \frac{\overline H_{\epsilon,\eta}\left(\rho_{AB}^{\mathrm{(sq)}}\right) -\max_{\rho_A}\overline H_{\epsilon,\eta}^{(1)}(\rho_{A}) }{\max_{\rho_A}\overline H_{\epsilon,\eta}^{(1)}(\rho_{A})},
\end{equation}
when the comparison is performed for fixed $n_A$ (see Fig.~\ref{fig: nA comparison}) or for fixed total number of photons $N$ (see Fig.~\ref{fig: N comparison}).
We see how two-mode squeezed states always yield a better average precision than all single-mode probes with the same $n_A$. Remarkably, in certain conditions the same remains true even if we compare states with the same total number of photons $N$, thus keeping into account also the photons in the ancillary mode. For different values of $\epsilon$ the qualitative behavior is retained, but the advantage $\mathcal I$ is reduced when $\epsilon$ is small (see Appendix \ref{app: epsilon plots} for the plots associated with $\epsilon = 0.5$ or $0.1$).  

Even in presence of losses, as in the noiseless case, the QFI associated with the correlated probe $\rho_{AB}^{\mathrm{(sq)}}$ does not depend on the direction of squeezing $\theta$ applied by the device under investigation. Hence, the stability of this particular class of entangled states against fluctuations in $\theta$ is not canceled by the introduction of photon losses.

\section{Conclusions} \label{sec: Conclusions}
In this paper we considered a black-box metrology problem in a Gaussian framework, where the goal is to estimate the squeezing power of a certain device in the absence of a-priori knowledge about the direction of application. We argued that it is reasonable to assume the knowledge of this phase at the measurement stage, after the chosen probe has been retrieved. We thus used the average QFI over different squeezing directions as a figure of merit to quantify the optimal performance of each probe. Indeed, we showed that this represents a natural choice not only if the squeezing direction is fixed but initially unknown, but also if it fluctuates randomly from one interaction to another.

In our analysis we analytically solved the problem for single-mode Gaussian inputs undergoing a noiseless evolution, showing that the optimal average performance can be obtained by using all the available energy to squeeze a vacuum state. However, the variance of the QFI associated with this probe necessarily increases with the photon number, potentially leading to the same precision obtainable with a vacuum input (i.e., QFI = 2) for the worst possible realization of $\theta$. In contrast, the same average precision can be obtained with no fluctuations in $\theta$ by employing a two-mode squeezed state with the same number of photons in the squeezed subsystem, at the price of doubling the total number of photons composing the probe.

We also numerically studied the same problem in presence of a noisy evolution, in which the transmission line leading to the squeezer is affected by photon losses. We showed that, in presence of loss, the choice of using all the available energy to squeeze the input probe might not be optimal, and better average performances could be obtained by introducing a displacement along the direction of the quadrature with minimal variance. Indeed, for high losses coherent states become the optimal single-mode probe. Once the strength $\epsilon$ of the squeezing device is roughly known, the dependence of this threshold on $n_A$ can be numerically computed as in Fig.~\ref{fig: optimal single-mode probe}. Finally, we numerically looked at the precision that could be reached by the paradigmatic example of two-mode squeezed states, in order to see if correlations could yield an advantage. We found that this is indeed the case, not only if the comparison is performed for fixed $n_A$, but, in some cases, also if we take into account the total photon number $N$ of the probe. Together with the independence of their QFI upon the squeezing direction, our results show how these states are good versatile probes, able to obtain an high estimation precision for all possible encoding realizations.

\section{Acknowledgements}
L.R. thanks M.S. Kim for valuable discussions, and acknowledges financial support from the People Programme (Marie Curie Actions) of the European Union’s Seventh Framework Programme (FP7/2007-2013) under REA Grant agreement 317232.
L.A.M.S. acknowledges the Brazilian agencies CAPES (Grant No. 6842/2014-03) and CNPq (Grant No. 470131/2013-6). G.A. acknowledges financial support from the European Research Council (ERC) through the StG GQCOP (Grant Agreement No. 637352), the Royal Society through the International Exchanges Programme (Grant No. IE150570), and the Foundational Questions Institute (fqxi.org) through the Physics of the Observer Programme (Grant No. FQXi-RFP-1601).

\bibliography{AvQFIbiblio}

\begin{thebibliography}{10}

\bibitem{Braunstein&Caves_1994}
Samuel~L. Braunstein and Carlton~M. Caves.
\newblock Statistical distance and the geometry of quantum states.
\newblock {\em Phys. Rev. Lett.}, 72:3439--3443, May 1994.

\bibitem{Giovannetti_2006}
Vittorio Giovannetti, Seth Lloyd, and Lorenzo Maccone.
\newblock Quantum metrology.
\newblock {\em Phys. Rev. Lett.}, 96:010401, Jan 2006.

\bibitem{Giovannetti_2011}
Vittorio Giovannetti, Seth Lloyd, and Lorenzo Maccone.
\newblock {Advances in quantum metrology}.
\newblock {\em Nat Photon}, 5(4):222--229, April 2011.

\bibitem{Dobrzanski_2012}
Rafal Demkowicz-Dobrzanski, Jan Kolodynski, and Madalin Guta.
\newblock {The elusive Heisenberg limit in quantum-enhanced metrology}.
\newblock {\em Nature Communications}, 3:1063+, September 2012.

\bibitem{Toth2014a}
Geza Toth and Iagoba Apellaniz.
\newblock Quantum metrology from a quantum information science perspective.
\newblock {\em Journal of Physics A: Mathematical and Theoretical},
  47(42):424006, 2014.

\bibitem{Dobrzanski_2014}
Rafal Demkowicz-Dobrza\ifmmode~\acute{n}\else \'{n}\fi{}ski and Lorenzo
  Maccone.
\newblock Using entanglement against noise in quantum metrology.
\newblock {\em Phys. Rev. Lett.}, 113:250801, Dec 2014.

\bibitem{Dobrzanski_2015}
Rafal Demkowicz-Dobrza\'{n}ski, Marcin Jarzyna, and Jan Ko\l{}ody\'{n}ski.
\newblock Chapter four - quantum limits in optical interferometry.
\newblock volume~60 of {\em Progress in Optics}, pages 345 -- 435. Elsevier,
  2015.

\bibitem{Schnabel_2010}
Roman Schnabel, Nergis Mavalvala, David~E. McClelland, and Ping~K. Lam.
\newblock {Quantum metrology for gravitational wave astronomy}.
\newblock {\em Nature Communications}, 1(8):121+, November 2010.

\bibitem{Dobrzanski_2013}
Rafa\l{} Demkowicz-Dobrza\ifmmode~\acute{n}\else \'{n}\fi{}ski, Konrad
  Banaszek, and Roman Schnabel.
\newblock Fundamental quantum interferometry bound for the
  squeezed-light-enhanced gravitational wave detector geo 600.
\newblock {\em Phys. Rev. A}, 88:041802, Oct 2013.

\bibitem{Taylor_2014}
Michael~A. Taylor and Warwick~P. Bowen.
\newblock Quantum metrology and its application in biology.
\newblock {\em Physics Reports}, 615:1 -- 59, 2016.
\newblock Quantum metrology and its application in biology.

\bibitem{Matthews_2016}
Jonathan~CF Matthews, Xiao-Qi Zhou, Hugo Cable, Peter~J Shadbolt, Dylan~J
  Saunders, Gabriel~A Durkin, Geoff~J Pryde, and Jeremy~L O’Brien.
\newblock Towards practical quantum metrology with photon counting.
\newblock {\em {NPJ Q}uantum Information}, 2:16023, August 2016.

\bibitem{Jachura_2016}
Micha{\l} Jachura, Rados{\l}aw Chrapkiewicz, Rafa{\l} Demkowicz-Dobrza\'{n}ski,
  Wojciech Wasilewski, and Konrad Banaszek.
\newblock {Mode engineering for realistic quantum-enhanced interferometry}.
\newblock {\em Nature Communications}, 7:11411, May 2016.

\bibitem{Girolami_2014}
Davide Girolami, Alexandre~M. Souza, Vittorio Giovannetti, Tommaso Tufarelli,
  Jefferson~G. Filgueiras, Roberto~S. Sarthour, Diogo~O. Soares-Pinto, Ivan~S.
  Oliveira, and Gerardo Adesso.
\newblock Quantum discord determines the interferometric power of quantum
  states.
\newblock {\em Phys. Rev. Lett.}, 112:210401, May 2014.

\bibitem{Nichols_16}
Rosanna Nichols, Thomas~R. Bromley, Luis~A. Correa, and Gerardo Adesso.
\newblock Practical quantum metrology in noisy environments.
\newblock {\em Phys. Rev. A}, 94:042101, Oct 2016.

\bibitem{Farace_2014}
Alessandro Farace, Antonella~De Pasquale, Luca Rigovacca, and Vittorio
  Giovannetti.
\newblock Discriminating strength: a bona fide measure of non-classical
  correlations.
\newblock {\em New Journal of Physics}, 16(7):073010, 2014.

\bibitem{Girolami_2013}
Davide Girolami, Tommaso Tufarelli, and Gerardo Adesso.
\newblock Characterizing nonclassical correlations via local quantum
  uncertainty.
\newblock {\em Phys. Rev. Lett.}, 110:240402, Jun 2013.

\bibitem{Roga_2014}
Wojciech Roga, Salvatore~M. Giampaolo, and Fabrizio Illuminati.
\newblock Discord of response.
\newblock {\em Journal of Physics A: Mathematical and Theoretical},
  47(36):365301, 2014.

\bibitem{Adesso_2016}
Gerardo Adesso, Thomas~R. Bromley, and Marco Cianciaruso.
\newblock Measures and applications of quantum correlations.
\newblock {\em Journal of Physics A: Mathematical and Theoretical},
  49(47):473001, 2016.

\bibitem{Roga_2016}
Wojciech Roga, Dominique Spehner, and Fabrizio Illuminati.
\newblock Geometric measures of quantum correlations: characterization,
  quantification, and comparison by distances and operations.
\newblock {\em Journal of Physics A: Mathematical and Theoretical},
  49(23):235301, 2016.

\bibitem{Bromley2016a}
Thomas~R. Bromley, Isabela~A. Silva, Charlie~O. Oncebay-Segura, Diogo~O.
  Soares-Pinto, Eduardo~R. deAzevedo, Tommaso Tufarelli, and Gerardo Adesso.
\newblock There is more to quantum interferometry than entanglement.
\newblock {\em arXiv:1610.07504}, 2016.

\bibitem{Braun2017a}
Daniel Braun, Gerardo Adesso, Fabio Benatti, Roberto Floreanini, Ugo Marzolino,
  Morgan~W. Mitchell, and Stefano Pirandola.
\newblock Quantum enhanced measurements without entanglement.
\newblock {\em arXiv:1701.05152}, 2017.

\bibitem{Ollivier_2001}
Harold Ollivier and Wojciech~H. Zurek.
\newblock Quantum discord: A measure of the quantumness of correlations.
\newblock {\em Phys. Rev. Lett.}, 88:017901, Dec 2001.

\bibitem{Henderson_2001}
Leah Henderson and Vlatko Vedral.
\newblock Classical, quantum and total correlations.
\newblock {\em Journal of Physics A: Mathematical and General}, 34(35):6899,
  2001.

\bibitem{Farace_2016}
Alessandro Farace, Antonella~De Pasquale, Gerardo Adesso, and Vittorio
  Giovannetti.
\newblock Building versatile bipartite probes for quantum metrology.
\newblock {\em New Journal of Physics}, 18(1):013049, 2016.

\bibitem{Ferraro_2005}
A.~Ferraro, S.~Olivares, and {Matteo G. A.} Paris.
\newblock {\em Gaussian States in Quantum Information}.
\newblock Napoli Series on physics and Astrophysics. Bibliopolis, 2005.

\bibitem{Weedbrook_2012}
Christian Weedbrook, Stefano Pirandola, Ra\'ul Garc\'{\i}a-Patr\'on, Nicolas~J.
  Cerf, Timothy~C. Ralph, Jeffrey~H. Shapiro, and Seth Lloyd.
\newblock Gaussian quantum information.
\newblock {\em Rev. Mod. Phys.}, 84:621--669, May 2012.

\bibitem{Adesso_2014_GaussReview}
Gerardo Adesso, Sammy Ragy, and Antony~R. Lee.
\newblock Continuous variable quantum information: Gaussian states and beyond.
\newblock {\em Open Systems and Information Dynamics}, 21(01n02):1440001, 2014.

\bibitem{Adesso_2014_GIP}
Gerardo Adesso.
\newblock Gaussian interferometric power.
\newblock {\em Phys. Rev. A}, 90:022321, Aug 2014.

\bibitem{Rigovacca_2015}
Luca Rigovacca, Alessandro Farace, Antonella De~Pasquale, and Vittorio
  Giovannetti.
\newblock Gaussian discriminating strength.
\newblock {\em Phys. Rev. A}, 92:042331, Oct 2015.

\bibitem{Roga_2015}
Wojciech Roga, Daniela Buono, and Fabrizio Illuminati.
\newblock Device-independent quantum reading and noise-assisted quantum
  transmitters.
\newblock {\em New Journal of Physics}, 17(1):013031, 2015.

\bibitem{Milburn_1994}
Gerard~J. Milburn, Wen-Yu Chen, and K.~R. Jones.
\newblock Hyperbolic phase and squeeze-parameter estimation.
\newblock {\em Phys. Rev. A}, 50:801--804, Jul 1994.

\bibitem{Chiribella_2006}
Giulio Chiribella, Giacomo~M. D'Ariano, and Massimiliano~F. Sacchi.
\newblock Optimal estimation of squeezing.
\newblock {\em Phys. Rev. A}, 73:062103, Jun 2006.

\bibitem{Paris_09}
Roberto Gaiba and Matteo~G.A. Paris.
\newblock Squeezed vacuum as a universal quantum probe.
\newblock {\em Physics Letters A}, 373(10):934 -- 939, 2009.

\bibitem{Safranek_2015}
Dominik \v{S}afr\'{a}nek, Antony~R Lee, and Ivette Fuentes.
\newblock Quantum parameter estimation using multi-mode gaussian states.
\newblock {\em New Journal of Physics}, 17(7):073016, 2015.

\bibitem{Safranek_16_OptGauss}
Dominik \ifmmode~\check{S}\else \v{S}\fi{}afr\'anek and Ivette Fuentes.
\newblock Optimal probe states for the estimation of gaussian unitary channels.
\newblock {\em Phys. Rev. A}, 94:062313, Dec 2016.

\bibitem{Genoni_2009}
Marco~G. Genoni, Carmen Invernizzi, and Matteo G.~A. Paris.
\newblock Enhancement of parameter estimation by kerr interaction.
\newblock {\em Phys. Rev. A}, 80:033842, Sep 2009.

\bibitem{Paris_2009}
Matteo G.~A. Paris.
\newblock Quantum estimation for quantum technology.
\newblock {\em International Journal of Quantum Information},
  07(supp01):125--137, 2009.

\bibitem{Pinel_2013}
Olivier Pinel, Pu~Jian, Nicolas Treps, Claude Fabre, and Daniel Braun.
\newblock Quantum parameter estimation using general single-mode gaussian
  states.
\newblock {\em Phys. Rev. A}, 88:040102, Oct 2013.

\bibitem{CramerBook}
Harald Cram\'{e}r.
\newblock {\em {Mathematical Methods of Statistics. (PMS-9)}}.
\newblock Princeton University Press, March 1999.

\bibitem{Nielsen&Chuang}
Michael~A. Nielsen and Isaac~L. Chuang.
\newblock {\em {Quantum Computation and Quantum Information}}.
\newblock Cambridge University Press, 1 edition, October 2004.

\bibitem{Bures_1969}
Donald Bures.
\newblock An extension of kakutani's theorem on infinite product measures to
  the tensor product of semifinite w*-algebras.
\newblock {\em Transactions of the American Mathematical Society},
  135:199--212, 1969.

\bibitem{Uhlmann_1976}
A.~Uhlmann.
\newblock The “transition probability” in the state space of a
  \textasteriskcentered-algebra.
\newblock {\em Reports on Mathematical Physics}, 9(2):273 -- 279, 1976.

\bibitem{Safranek_2016}
Dominik {{\v S}afr{\'a}nek}.
\newblock {Discontinuities of the quantum Fisher information and the Bures
  metric}.
\newblock {\em ArXiv: 1612.04581}, December 2016.

\bibitem{paramestimPRA}
Antonella~De Pasquale, Davide Rossini, Paolo Facchi, and Vittorio Giovannetti.
\newblock Quantum parameter estimation affected by unitary disturbance.
\newblock {\em Phys. Rev. A}, 88:052117, Nov 2013.

\bibitem{Genoni_2016}
Marco~G. Genoni, Ludovico Lami, and Alessio Serafini.
\newblock Conditional and unconditional gaussian quantum dynamics.
\newblock {\em Contemporary Physics}, 57(3):331--349, 2016.

\bibitem{Banchi_2015}
Leonardo Banchi, Samuel~L. Braunstein, and Stefano Pirandola.
\newblock Quantum fidelity for arbitrary gaussian states.
\newblock {\em Phys. Rev. Lett.}, 115:260501, Dec 2015.

\bibitem{Friis_15}
Nicolai Friis, Michalis Skotiniotis, Ivette Fuentes, and Wolfgang D\"ur.
\newblock Heisenberg scaling in gaussian quantum metrology.
\newblock {\em Phys. Rev. A}, 92:022106, Aug 2015.

\bibitem{Wigner_SkewInfo_1963}
Eugene~P. Wigner and Mutsuo~M. Yanase.
\newblock Information contents of distributions.
\newblock {\em Proceedings of the National Academy of Sciences},
  49(6):910--918, 1963.

\bibitem{Luo_SkewInfo_2003}
Shunlong Luo.
\newblock Wigner-yanase skew information and uncertainty relations.
\newblock {\em Phys. Rev. Lett.}, 91:180403, Oct 2003.

\bibitem{Lupo_2012}
Cosmo Lupo, Stefano Mancini, Antonella~De Pasquale, Paolo Facchi, Giuseppe
  Florio, and Saverio Pascazio.
\newblock Invariant measures on multimode quantum gaussian states.
\newblock {\em Journal of Mathematical Physics}, 53(12):122209, 2012.

\bibitem{Zyczkowski_2011}
Karol \.{Z}yczkowski, Karol~A. Penson, Ion Nechita, and Beno\^{\i}t Collins.
\newblock Generating random density matrices.
\newblock {\em Journal of Mathematical Physics}, 52(6):062201, 2011.

\end{thebibliography}
\bibliographystyle{unsrt}

\clearpage
\appendix
\onecolumngrid

\section{AvQFI characterizes the ultimate estimation precision for a fluctuating interaction} \label{app: AvQFI fluctuating theta}

We remind the reader that, in a black-box metrology setting, the parameter $\theta$ characterizing the interactions between the probes and the system of interest is typically not allowed to change once it has been randomly picked and communicated to the experimenter. 
In the model described in Fig.\ref{fig: sketch}, this could correspond to a fixed, but initially unknown, optical path length separating the probes from the squeezer.
However, one could imagine a different scenario, in which the probe-system interaction characterizing the evolution can change randomly from one probe to the next, and all choices are then communicated to the experimenter. With the same physical model in mind, this could correspond to a fluctuating optical path length that, for example, could be deduced \emph{a posteriori} from the travel time of the probe. As before, this knowledge allows the experimenter to perform the optimal measurement on each of the $M$ recollected states of the probes.
In  this situation the RMSE  of any  estimator $\hat{\epsilon}$ for the parameter $\epsilon$ can be lower bounded by 
\begin{equation}\label{NEWEQ2}
 \delta {\hat{\epsilon}} \geq \frac{1}{\sqrt{M \overline H_{\epsilon}(\rho)}},
\end{equation}
where $M$ is the number of times the experiment is repeated.
This result shows that the AvQFI also characterizes the ultimate precision bound attainable by a given probe in the physically relevant regime where the details of the evolution cannot be controlled, but only detected \emph{a posteriori}. When this is the case, the only viable strategy is to look at versatile probes, characterized by high AvQFI values.

In order to derive Eq.~(\ref{NEWEQ2}) we consider a situation in which all possible probe-system interactions are labeled by an unknown parameter $\theta$, that is also free to fluctuate from one interaction to another according to some probability $p(\theta)$. Every interaction also depends on a real parameter $\epsilon$, that we want to estimate.
In the specific example discussed in this paper,  $\theta$ and $\epsilon$ represent respectively the squeezing direction and strength, while $p(\theta)$ is taken to be uniform in $[0, 2\pi]$.  
In what follows, we show that the weighted average of the $\theta$-dependent QFIs characterizes the ultimate estimation precision, as long as the specific realizations of $\theta$ are known at the measurement stage.

Upon recollection of a probe, we are communicated the parameter $\theta$ that affected its evolution, and we can exploit this information to choose a suitable POVM $\{E^{(\theta)}_x\}_x$. This measurement on the encoded state $\rho_{\epsilon,\theta}$ yields outcome $x$ with probability $p(x|\epsilon,\theta) = \Tr{\rho_{\epsilon,\theta} E^{(\theta)}_x}$. Once these measurements have been performed on each of the $M$ probes, we are left with a classical problem in which $\epsilon$ needs to be estimated from the knowledge of $M$ pairs $(x,\theta)$, sampled according to the distribution 
\begin{equation}
p(x,\theta|\epsilon) = p(x|\epsilon,\theta) p(\theta).
\end{equation}
The variance of any unbiased estimator $\hat{\epsilon}$ can thus be bounded via the classical Cram\'{e}r-Rao bound [see \Eq{eq: classical CRB} and \Eq{def: Fisher info}] associated with this probability distribution. In particular, the classical Fisher information of $p(x,\theta|\epsilon)$ is given by:
\begin{equation}
F_\epsilon = \int \text{d}x \,\text{d}\theta \, p(x,\theta|\epsilon) \left[\partial_\epsilon \ln p(x,\theta|\epsilon)\right]^2 = \int \text{d}\theta \, p(\theta) F_{\epsilon}^{(\theta)},
\end{equation}
where $F_{\epsilon}^{(\theta)}$ is the classical Fisher information of the probability distribution $p(x|\epsilon,\theta)$ obtained for fixed $\theta$. Therefore, for any choice of POVMs this reasoning yields the bound:
\begin{equation}
 \delta {\hat{\epsilon}} \geq \frac{1}{\sqrt{M \int \text{d}\theta \, p(\theta) F_{\epsilon}^{(\theta)}}}.
\end{equation}
Finally, notice that the right-hand side can be minimized by suitably choosing for every $\theta$ the POVM maximizing the Fisher information $F_{\epsilon}^{(\theta)}$. Since this is exactly the optimization that defines the quantum Fisher information $H_\epsilon^{(\theta)}(\rho)$ of the probe $\rho$, 
we are left with the following ultimate bound on the variance of any unbiased estimator:
\begin{equation}
 \delta {\hat{\epsilon}} \geq \frac{1}{\sqrt{M \overline H_\epsilon(\rho)}},
\end{equation}
where we defined the average QFI as
\begin{equation}
\overline H_\epsilon(\rho) = \int \text{d}\theta \, p(\theta) H_{\epsilon}^{(\theta)}(\rho),
\end{equation}
by taking into account the possibility of dealing with a non-uniform distribution $p(\theta)$.

\section{Standard form of covariance matrices}
\label{AppStd}
We can decompose the covariance matrix and the displacement vector of a two-mode probe state in the following blocks:
\begin{equation}
\Gamma = \left(\begin{array}{c|c}
\Gamma_A & \Gamma_{OFF}\\\hline
\Gamma_{OFF}^\intercal & \Gamma_B
\end{array}\right), \qquad \bxi = (\bxi_A,\bxi_B)^\intercal.
\end{equation}
By applying the Williamson decomposition on the $B$ subsystem, one can find a local Gaussian unitary map $\mathcal T^{(B)}$ which changes
\begin{equation}
\Gamma_B \to T^{-1}\Gamma_B T^{-1\intercal} = b \Id_2,\quad \text{with}\quad b\geq 1,
\end{equation}
and
\begin{equation}
\Gamma_{OFF} \to \Gamma_{OFF}T^{-1\intercal}, \qquad  \bxi_B \to \bz,
\end{equation}
while leaving subsystem $A$ unchanged. We stress that $b$ is the symplectic eigenvalue of the reduced covariance matrix $\Gamma_B$, and not one of the symplectic eigenvalues of the total matrix $\Gamma$. At this stage, thanks to the singular value decomposition (SVD), one can write 
\begin{equation}
\Gamma_{OFF}T^{-1\intercal} = R_1 \text{Diag}(c,d) R_2^\intercal,
\end{equation}
for some $R_1, R_2 \in SO(2)$ and real (not necessarily positive) parameters $c,d$. 
If $\mathcal R_1^{(A)},\mathcal R_1^{(B)}$ are the single-mode rotations respectively associated with $R_1^{-1},R_2^{-1}$ via \Eq{eq: CV unitary change}, the overall Gaussian unitary $\mathcal R_1^{(A)}\otimes \mathcal R_1^{(B)} \circ \mathcal T^{(B)}$ transforms the input probe to an equivalent one that we label $\rho^{\text{(std)}}$. $\rho^{\text{(std)}}$ has the same average QFI of the original $\rho$, but the simpler structure:
\begin{equation}\label{def: std parametrization app}
\Gamma^{\text{(std)}} = \left(\begin{array}{cc|cc}
a_x & a_{xp} & c & 0 \\
a_{xp} & a_p & 0 & d\\\hline
c & 0 & b & 0 \\
0 & d & 0 & b
\end{array}\right), \quad \bxi^{\text{(std)}} = (\xi_x,\xi_p,0,0)^\intercal,
\end{equation}
and we can limit our analysis to states of this form.

For the reader familiar with the topic of quantum correlations in Gaussian states, we stress that this standard form is different from the one typically used when discussing Gaussian entanglement because the AvQFI is not invariant under generic Gaussian unitary operations on A.

\section{Noiseless two-mode AvQFI}\label{app: Noiseless two-mode AvQFI}
We report here the general expression for the noiseless two-mode AvQFI for squeezing estimation, as function of the parameters appearing in the standard form of \Eq{def: std parametrization}. The contribution $\overline{H}_{\text{disp}}$ coming from the displacement is written separately.
\begin{equation}\label{eq: two-mode AvQFI displacement}
\overline{H}_{\text{disp}}[\rho_{AB}] = \frac{b |\xi|^2}{\det\Gamma}\left(b(a_x + a_p) -c^2 - d^2\right).
\end{equation}
\begin{align}
\overline {H} [\rho_{AB}] & =\overline{H}_{\text{disp}}[\rho_{AB}] \plus \frac{c^2 \left(4 d^2 \minus b (a_x +5 a_p )\right)+b \left(b \left(a_x ^2+6 a_x  a_p +a_p ^2 \minus 4 a_{xp} ^2\right) \minus d^2 (5 a_x +a_p )\right)}{2 \left(-b^2 a_{xp} ^2+\left(c^2-a_x  b\right) \left(d^2-b a_p \right)-1\right)}\notag\\
&\hspace*{-1cm} \minus \frac{4 \left(b^2 \plus c d \plus 1\right) \left( \minus \left(b^2 \plus 1\right) a_{xp} ^2 \plus a_x  \left(a_p  b^2 \minus b d^2 \plus a_p \right) \plus c^2 \left(d^2 \minus b a_p \right) \plus c d\right) \plus \left(a_x \plus b^2 (a_x \plus a_p ) \minus b \left(c^2 \plus d^2\right) \plus a_p \right)^2}{2 \left(-b^2 a_{xp} ^2 + \left(c^2-a_x  b\right) \left(d^2-b a_p \right)-1\right) \left(a_x  a_p +b^2 \left(a_x  a_p -a_{xp} ^2+1\right)-b \left(a_p  c^2+a_x  d^2\right)-a_{xp} ^2+(c d+1)^2\right)}.
\end{align}

\section{Single-mode probes with maximum and minimum AvQFI}\label{app: noiseless bounds}
Among all single-mode probes with the same photon-number $n_A$, pure squeezed states maximize the AvQFI of \Eq{eq: SM AvQFI}, while thermal states minimize it. In this appendix we will formally prove this statement. Notice that if the input covariance matrix is parametrized as in \Eq{eq: single-mode Gauss state}, the AvQFI is a function of the absolute value of the input displacement $|\xi|$ and of the quantities:
\begin{equation}
\Tr{\Gamma_A} = 2 \nu_A \cosh(2\alpha), \qquad \det\Gamma_A = \nu_A^2.
\end{equation}

Let us start with the maximum. By exploiting the expression for $n_A$ given in \Eq{def: formula n_A}, we can substitute the parameter $\alpha$ in $\overline H$ and obtain
\begin{equation}
\overline H|_{n_A} = 2 \frac{(2 n_A +1 - |\xi|^2)^2 + \nu^2}{1+ \nu^2} + 2 |\xi|^2\frac{2 n_A +1 - |\xi|^2}{\nu^2}.
\end{equation}
With a simple algebra, it is easy to see that the displacement has an overall negative contribution in the AvQFI expression. Therefore, if the photon-number is fixed, reducing the squeezing always improves the performance of the probe. We are thus left with the task of showing that $\overline H|_{n_A}$ is maximized by $\nu \leq 1$ when $|\xi|=0$. To do so, notice that the first derivative of the function
\begin{equation}
f_{a,b}(x) = \frac{a + x}{b + x},
\end{equation}
has the same sign of $b-a$. This concludes the proof because the average QFI can be written as
\begin{equation}
\overline H|_{n_A, |\xi|=0} = 2 f_{(2 n_A +1)^2, 1}(\nu),
\end{equation}
and is  thus maximized by choosing the minimum symplectic eigenvalue $\nu=1$.

We now turn to the problem of finding the probe which minimizes $\overline H$ for fixed $n_A$. By using \Eq{def: formula n_A} to substitute the value of $\nu$ in the formula for the average QFI given in \Eq{eq: SM AvQFI}, we find
\begin{align}
\overline H|_{n_A} = 2 (2 n_A +1 - |\xi|^2)^2\frac{1+\cosh^2(2\alpha) }{(2 n_A +1 - |\xi|^2)^2 + \cosh^2(2\alpha)}  +2 |\xi|^2 \cosh^2(2\alpha) \frac{1}{(2 n_A +1 - |\xi|)}.
\end{align}
The first fraction appearing in this expression can be written as $f_{a,b}(\cosh^2[2\alpha])$ for 
$a = 1$ and $b = (2 n_A +1 - |\xi|^2)^2 \geq 1$.
The average QFI is then monotonically increasing with $\cosh(2\alpha)$, and is minimized when $\alpha =0$. Finally, we have to show that the minimum $\overline H|_{n_A,\alpha = 0}$ is reached for undisplaced probes. We do this by showing that its first derivative on $|\xi|^2$ is always positive. After some straightforward manipulations, this inequality can be written as
\begin{equation}
\sqrt{2}\left(Y- 2 X\right)(Y+ 2 X) \geq 0,
\end{equation}
where we introduced the auxiliary positive quantities
\begin{align}
X = (2n_A + 1 - |\xi|^2), \quad
Y = \sqrt{2 n_A +1}\left[1+X^2\right].
\end{align}
The proof is now concluded because
\begin{equation}
Y- 2X = \left(\sqrt{2n_A +1} -1\right)(1+ X^2) + (1-X)^2 \geq 0.
\end{equation}

\section{Uniform sampling of single-mode Gaussian states}\label{app: sampling appendix}
In this appendix we review the method used in Sec.~\ref{sec: noisy} to sample single-mode Gaussian states with fixed number of photons, uniformly distributed according to the unique invariant measure induced by the left Haar measure on the group of Gaussian unitaries. Further details on multimode generalizations can be found in Ref. \cite{Lupo_2012} or references therein.

Let us start with pure single-mode Gaussian states, which can be written as
\begin{equation}\label{eq: single-mode pure Gaussian}
\ketbra{\psi_G^{(1)}}{\psi_G^{(1)}} = \mathcal U_G^{(A)} \left[\ketbra{0}{0}\right],
\end{equation}
where $\ket{0}$ represents the vacuum state and $\mathcal U_G^{(A)}$ is a generic single-mode Gaussianity preserving unitary map. For a single-mode system we can write it as
\begin{equation}
\mathcal U_G^{(A)} = \mathcal D^{(A)}_{\bxi_A}\circ \mathcal R^{(A)}_\phi\circ \mathcal S_\alpha^{(A)} \circ\mathcal R_{\phi^\prime}^{(A)},
\end{equation}
where $\mathcal D_{\bxi_A}^{(A)}$ is the displacement map which acts on the quadratures $\rop_A$ as
\begin{equation}
\rop_A  \stackrel{\mathcal D^{(A)}_{\bxi_A}}{\longrightarrow} \rop_A + \bxi_A.
\end{equation}
With this parametrization, and using the following polar decomposition in phase space
\begin{equation}
\bxi_A = (|\xi|\cos\psi, |\xi|\sin\psi)^\intercal 
\end{equation}
the Haar invariant measure on the group of $1$-mode Gaussian unitaries is given by
\begin{equation}\label{eq: single-mode unitary invariant measure}
\text{d} (\mathcal U_G^{(A)}) = \frac{\mathcal N_1}{2} \;\text{d} (\cosh \alpha) \,\text{d}\phi \,\text{d}\phi^\prime \,\text{d}\left(|\xi|^2\right) \,\text{d}\psi,
\end{equation}
up to a normalization constant $\mathcal N_1$. In order to impose an energy constraint, say of $n_A$ photons, we can use the parametrization for $n_A$ given in \Eq{def: formula n_A} and add the Dirac delta function
\begin{equation}
\delta \left(n_A -\frac{\cosh(2 \alpha) -1 + |\xi^2|}{2}\right).
\end{equation}
This induces an invariant measure on the set of single-mode pure Gaussian states with fixed number of photons via \Eq{eq: single-mode pure Gaussian}:
\begin{equation}\label{def: single-mode invariant measure}
\text{d} \left(\psi_G^{(1)}\right)|_{n_A} = \mathcal N_1 \;\text{d} (\cosh \alpha) \,\text{d}\phi \, \text{d}\psi,
\end{equation}
with the constraint 
\begin{equation}\label{eq: displacement from nA}
|\xi|^2 = 2n_A +1 - \cosh(2\alpha).
\end{equation}
Notice that the dependence on $\phi^\prime$ disappears because it has no effect when $\mathcal U_G^{(A)}$ acts on the vacuum. 

If we had to sample a mixed single-mode Gaussian state, we can first sample a pure two-mode state, with an approach similar to the one just described, and then apply a partial trace over one of the two subsystems. This is a standard approach in generating random mixed quantum states (see e.g., Ref. \cite{Zyczkowski_2011}).
Following Ref.~\cite{Lupo_2012}, a pure two-mode Gaussian state can be written as
\begin{equation}
\ketbra{\psi_G^{(2)}}{\psi_G^{(2)}} = \mathcal{U}_G^{(A)}\otimes \mathcal{U}_G^{(B)} \left[ \ketbra{TMSV (\nu)}{TMSV (\nu)} \right],
\end{equation}
where $\ket{TMSV(\nu)}$ is a two-mode squeezed vacuum state
\begin{equation}
\ket{TMSV (\nu)} = \sqrt{\frac{2}{\nu+1}} \sum_{j=0}^\infty \left(\frac{\nu-1}{\nu+1}\right)^{j/2}\ket{j,j}.
\end{equation}
Note that the parameter $\nu$ corresponds to the symplectic eigenvalue of the reduced single-mode state obtained by tracing away the second mode.
The invariant measure on the manifold of pure two-mode Gaussian states then is
\begin{equation}
	\text{d}\left(\psi_G^{(2)}\right) = \frac{\mathcal N_2}{3} \, \text{d}(\nu^3) \, \text{d}\mathcal U_G^{(A)} \, \text{d}\mathcal U_G^{(B)} .
\end{equation}
If mode $B$ is traced away, the invariant measure for a mixed single-mode Gaussian state becomes
\begin{equation}
\text{d} \left(\rho_G^{(1)}\right) = \frac{\mathcal{N}_2 \mathcal{N}_1}{6} \text{d}(\nu^3) \, \text{d} (\cosh \alpha) \,\text{d}\left(|\xi|^2\right) \, \text{d}\phi \,\text{d}\psi,
\end{equation}
where we used \Eq{eq: single-mode unitary invariant measure} without the irrelevant angle $\phi^\prime$ (it has no effect on a thermal state).
Therefore, in order to uniformly sample a mixed single-mode Gaussian state with respect to this invariant measure, we need to: (i) uniformly sample $\nu^3, \cosh(\alpha)$, and $|\xi|^2$ within the region of $\mathbb R^3$ allowed by the energy constraints, and (ii) uniformly sample $\phi,\psi$ in $[0,2\pi]$.

\section{Single-mode noisy QFI and optimal displacement direction} \label{app: expressions noisy QFI}
By setting without loss of generality $\phi = 0$ in \Eq{eq: single-mode Gauss state}, the single-mode $\theta$-dependent QFI in the noisy case can be written as:
\begin{equation}
H^{(1)}_{\epsilon,\theta} = \frac{N_1}{D_1} + \frac{N_2}{D_2} + \frac{N_3}{D_3},
\end{equation}
where the coefficients $N_1, N_2, N_3, D_1, D_2, D_3$ are defined as:

\begin{align}
& N_1 = (\eta -1)^2 \eta ^2 e^{-4 (\alpha +\epsilon )} \left(\left(e^{4 \alpha }-1\right) \eta  \nu  \left(e^{4 \epsilon }+1\right) \cos (2 \theta )+4 e^{2 (\alpha +\epsilon )} \sinh (2 \epsilon ) (\eta  \nu  \cosh (2 \alpha )-\eta +1)\right)^2, \\
& N_2 = e^{-2 \alpha } \eta ^2 \left(\left(e^{4 \alpha }-1\right)^2 \eta ^2 \nu ^2 \cos (4 \theta )-2 e^{4 \alpha } \left(\eta  \nu  (\eta  \nu  \cosh (4 \alpha )-8 (\eta -1) \cosh (2 \alpha ))+\eta  \left(\eta  \left(3 \nu ^2+4\right)-8\right)+4\right)\right), \\
& \frac{N_3}{\eta ^2} = -\eta  \left(-2 \left(e^{4 \alpha }-1\right) \eta  \nu  \sin (4 \theta ) \xi _p \xi _x+\left(e^{4 \alpha }-1\right) \eta  \nu  \cos (4 \theta ) \left(\xi _p-\xi _x\right) \left(\xi _p+\xi _x\right)+2 e^{2 \alpha } \left(\xi _p^2+\xi _x^2\right) (\eta  \nu  \cosh (2 \alpha )-\eta +1)\right) \notag \\
& \qquad +2 e^{2 \alpha } (\eta -1) \sinh (2 \epsilon ) \left(-\cos (2 \theta ) \xi _p^2+2 \sin (2 \theta ) \xi _p \xi _x+\cos (2 \theta ) \xi _x^2\right)+2 e^{2 \alpha } (\eta -1) \cosh (2 \epsilon ) \left(\xi _p^2+\xi _x^2\right), \label{appeq: num disp} \\
& \frac{D_2}{2} = e^{4 \alpha } (\eta -1) \eta ^2 \nu  (\eta +\cos (2 \theta ) \sinh (2 \epsilon )+\cosh (2 \epsilon ))-e^{2 \alpha } \left(\eta  \left(\eta  \left(\eta  \left(\eta  \nu ^2+\eta -2\right)+2\right)-2\right)+2 \eta  (\eta -1)^2 \cosh (2 \epsilon )+2\right)\notag \\
& \qquad +(\eta -1) \eta ^2 \nu  (\eta -\cos (2 \theta ) \sinh (2 \epsilon )+\cosh (2 \epsilon)), \\
& \frac{D_3}{e^{2 \alpha }} =  2 (\eta -1) \eta  (\eta  \nu  \sinh (2 \alpha ) \cos (2 \theta ) \sinh (2 \epsilon )+\eta  \nu  \cosh (2 \alpha ) (\eta +\cosh (2 \epsilon ))-(\eta -1) \cosh (2 \epsilon )) \notag \\
& \qquad  +\eta ^4 \left(-\nu ^2\right)-(\eta -1)^2 \left(\eta ^2+1\right), \label{appeq: den disp}\\ 
&\sqrt{4 e^{4 (\alpha +\epsilon)}\left(\frac{D_1}{2} +1\right)} =  \left(e^{4 \alpha}-1\right) (1-\eta) \eta^2 \nu  \left(e^{4 \epsilon }-1\right) \cos (2 \theta ) \notag\\ 
&\qquad  +2 e^{2 (\alpha +\epsilon )} \left[2 (1-\eta) \eta  (\eta  \nu  \cosh (2 \alpha ) (\eta +\cosh (2 \epsilon ))+(1-\eta) \cosh (2 \epsilon ))+\eta ^4 \nu ^2+(1-\eta)^2 \left(\eta ^2+1\right)\right].
\end{align}

In particular, note that only the term $N_3/D_3$ depends on the displacement vector $\left(\xi_x,\xi_p\right)$ of the probe, characterized by the components $\xi_x = |\xi|\cos\psi$ and $\xi_p = |\xi| \sin\psi$. It is possible to show that the optimal displacement direction, leading to the largest AvQFI value, is $\psi = \pi/2$. In the remainder of this appendix we provide a formal proof of this statement.
First, we explicitly show the dependence of $N_3/D_3$ on the angles $\theta$ and $\psi$ by rewriting it as
\begin{equation}\label{appeq: disp term}
\frac{N_3}{D_3} = \frac{x_0 + x_1 \cos[2(\theta + \psi)] + x_2 \cos[2(2\theta + \psi)]}{x_3 + x_4\cos(2\theta)},
\end{equation} 
where the coefficients $\{x_i\}_{i=0}^4$ depend on $\epsilon,\eta,\alpha $ and $\nu$; 
in particular 
\begin{equation}
x_4 = \nu \eta^2 (1-\eta)(e^{4\alpha}-1)\sinh(2\epsilon) \geq 0.
\end{equation}
When $x_4 = 0$ the dependence upon $\psi$ disappears when we take the average over $\theta$. If $x_4 \neq 0$, by assuming without loss of generality $\epsilon,\alpha \geq 0$, note that a comparison with Eqs.~\eqref{appeq: num disp} and \eqref{appeq: den disp} yields the inequalities:
\begin{equation}\label{appeq: inequalities}
x_1 \geq 0, \qquad x_2 \leq 0, \qquad x_3/x_4 > 1.
\end{equation} 
Then, we explicitly perform the integration  of \Eq{appeq: disp term} over $\theta \in [0,2\pi]$. By dividing the result by $2 \pi$, we obtain the following contribution to the AvQFI:

\begin{equation}
\int_{0}^{2\pi} \frac{N_3}{D_3}\frac{\textit{d}\theta}{2\pi} =
\frac{x_0}{2 \sqrt{x_3^2 - x_4^2}}
+\frac{\cos 2\psi}{2 x_4}\left[ x_1 \int_{0}^{2\pi}\frac{\textit{d}\theta}{2\pi} \frac{\cos \theta}{\frac{x_3}{x_4} + \cos \theta} 
+ x_2 \int_{0}^{2\pi}\frac{\textit{d}\theta}{2\pi} \frac{\cos 2 \theta}{\frac{x_3}{x_4} + \cos \theta}\right],
\end{equation}

whose maximum value is reached when $\psi = \pm \pi/2$, because the term between square brackets is negative. This follows from   \Eq{appeq: inequalities}, and from the two inequalities:
\begin{equation}
\int_{0}^{2\pi}\frac{\textit{d}\theta}{2\pi} \frac{\cos \theta}{\frac{x_3}{x_4} + \cos \theta} \leq 0,\qquad \int_{0}^{2\pi}\frac{\textit{d}\theta}{2\pi} \frac{\cos 2 \theta}{\frac{x_3}{x_4} + \cos \theta} \geq 0,
\end{equation}
which hold for $x_3/x_4 > 1$.

\section{Dependence upon $\epsilon$}
\label{app: epsilon plots}
\begin{figure}[hbt]
		\centering
		\subfloat[$\epsilon = 0.5$. ]{
			\includegraphics[scale=0.7]{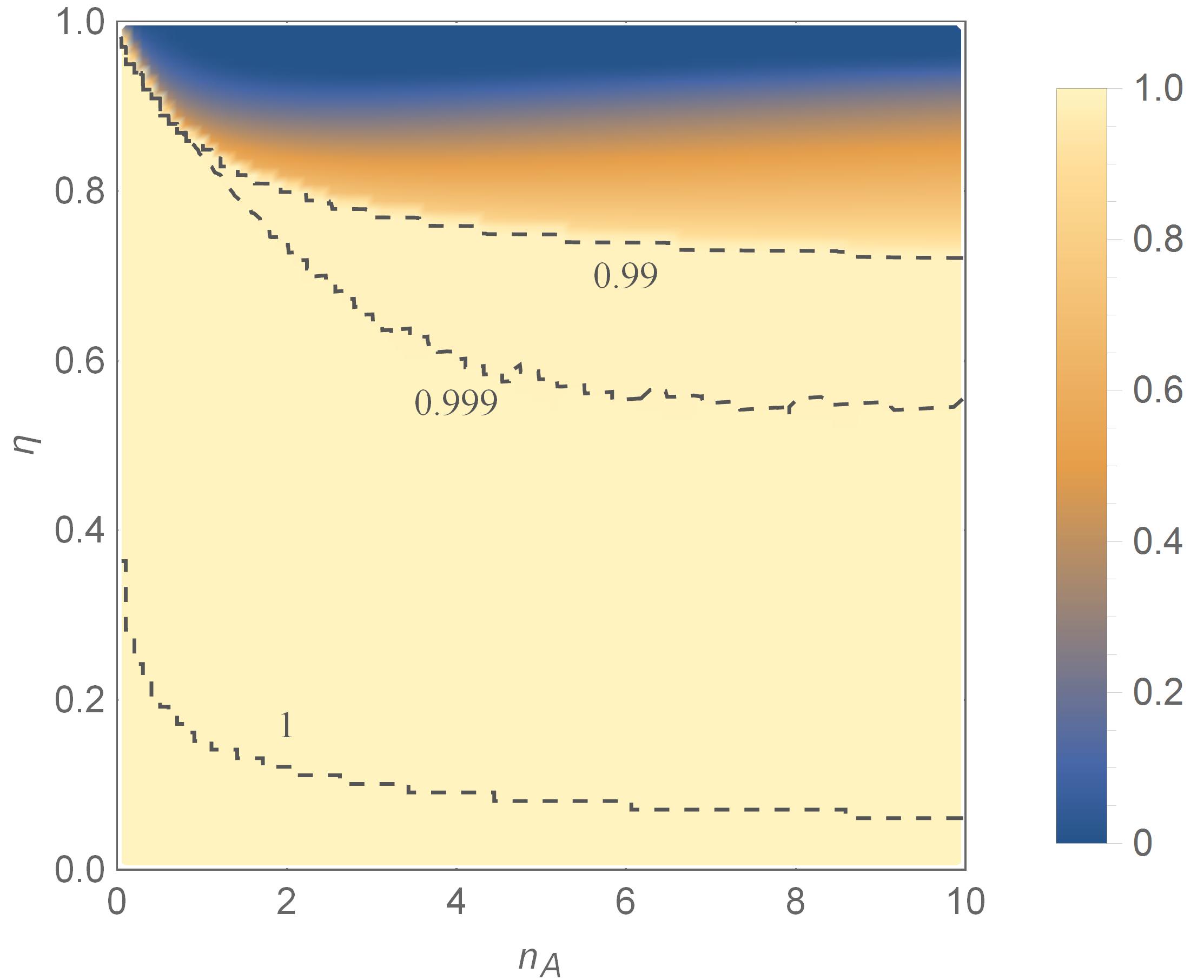}}
		\subfloat[$\epsilon = 0.1$.]{
			\includegraphics[scale=0.7]{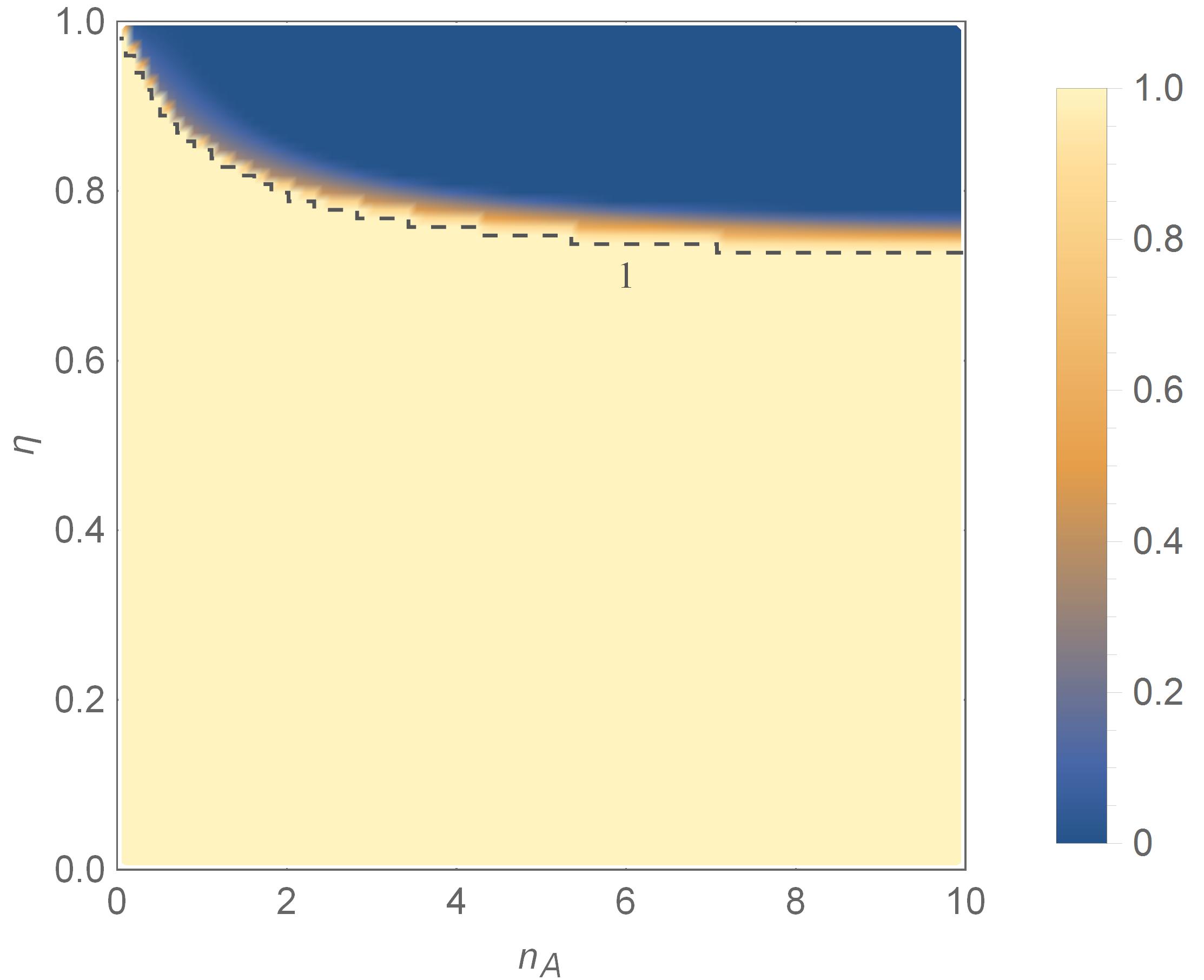}}
			\caption{(Color online)  Optimal value for the ratio $|\xi|^2/(2 n_A)$, leading to the maximum AvQFI value for $10^4$ pairs $(n_A,\eta) \in [0,10]\times[0,1]$ and different values of $\epsilon$. \label{fig: optimal states} }
	\end{figure}

\begin{figure}
		\centering
		\subfloat[$\epsilon = 0.5$. ]{
			\includegraphics[scale=0.7]{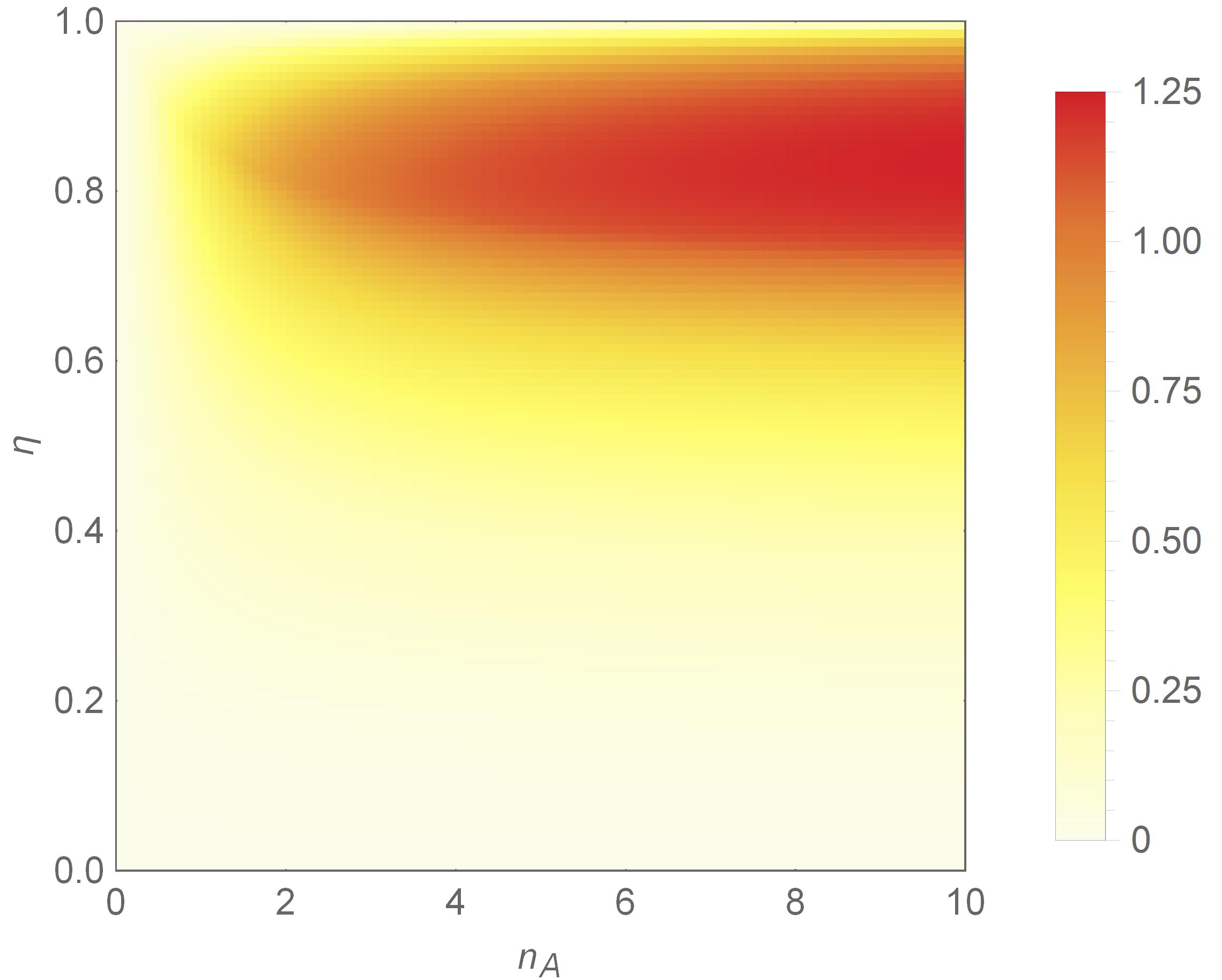}}
		\subfloat[$\epsilon = 0.1$.]{
			\includegraphics[scale=0.7]{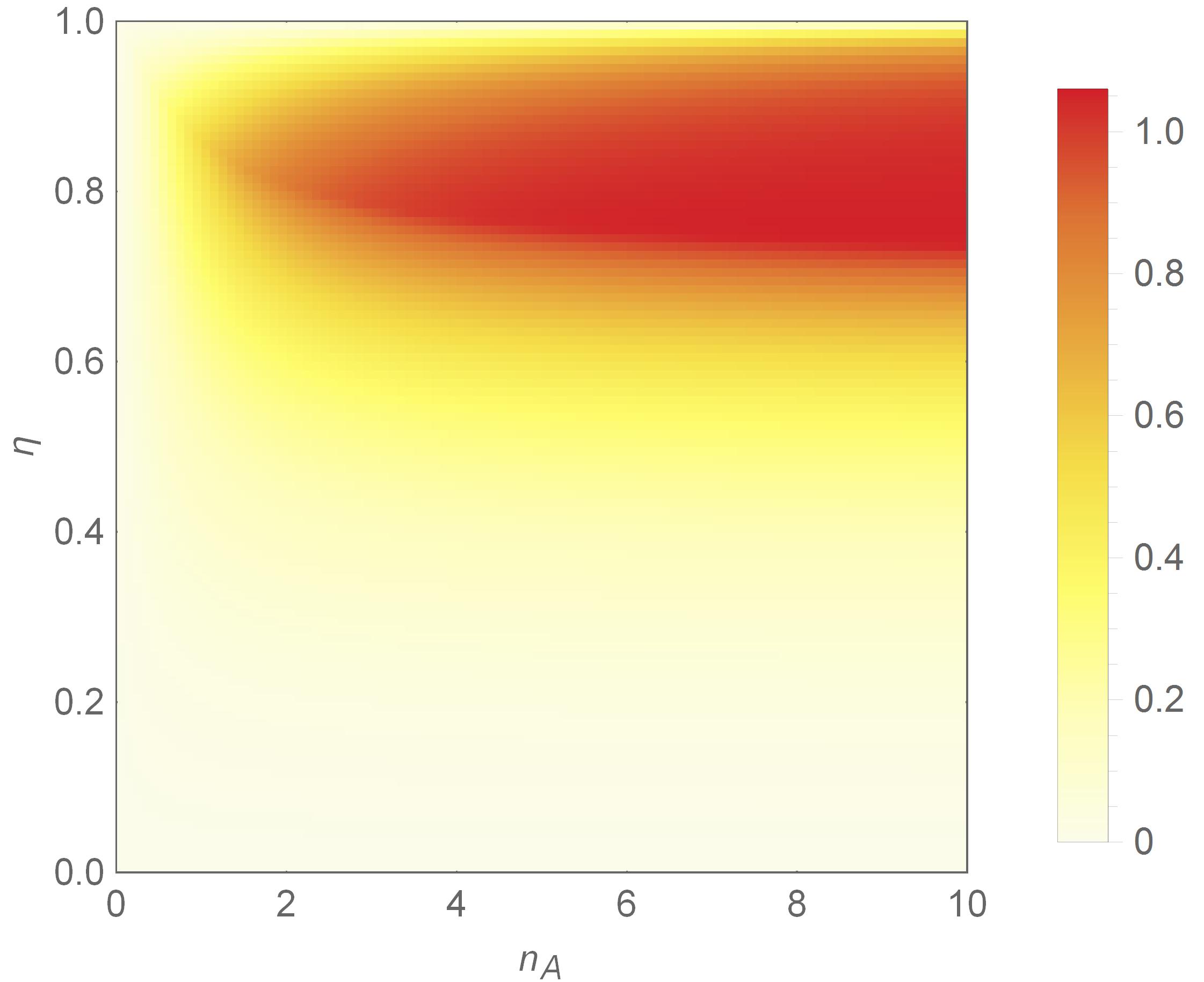}}
		\caption{(Color online) Relative increase $\mathcal I$ in precision for fixed $n_A$ obtained by using two-mode squeezed vacuum probes rather than the optimal single-mode input state, numerically evaluated for $10^4$ pairs $(n_A, \eta)\in [0,10]\times[0,1]$ and different values of $\epsilon$. \label{fig: increase vs nA}}
	\end{figure}

\begin{figure}
		\centering
		\subfloat[$\epsilon = 0.5$. ]{
			\includegraphics[scale=0.7]{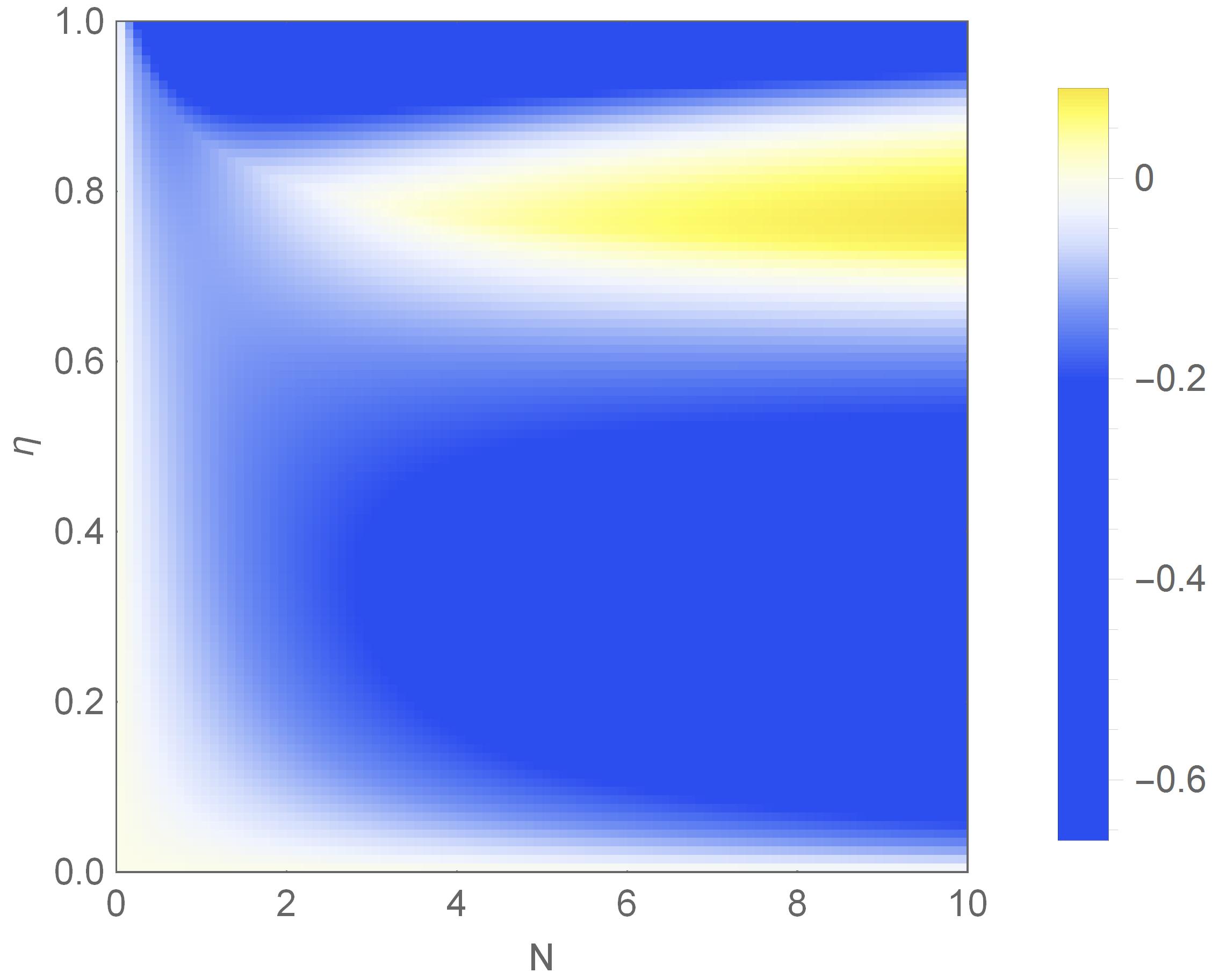}}
		\subfloat[$\epsilon = 0.1$.]{
			\includegraphics[scale=0.7]{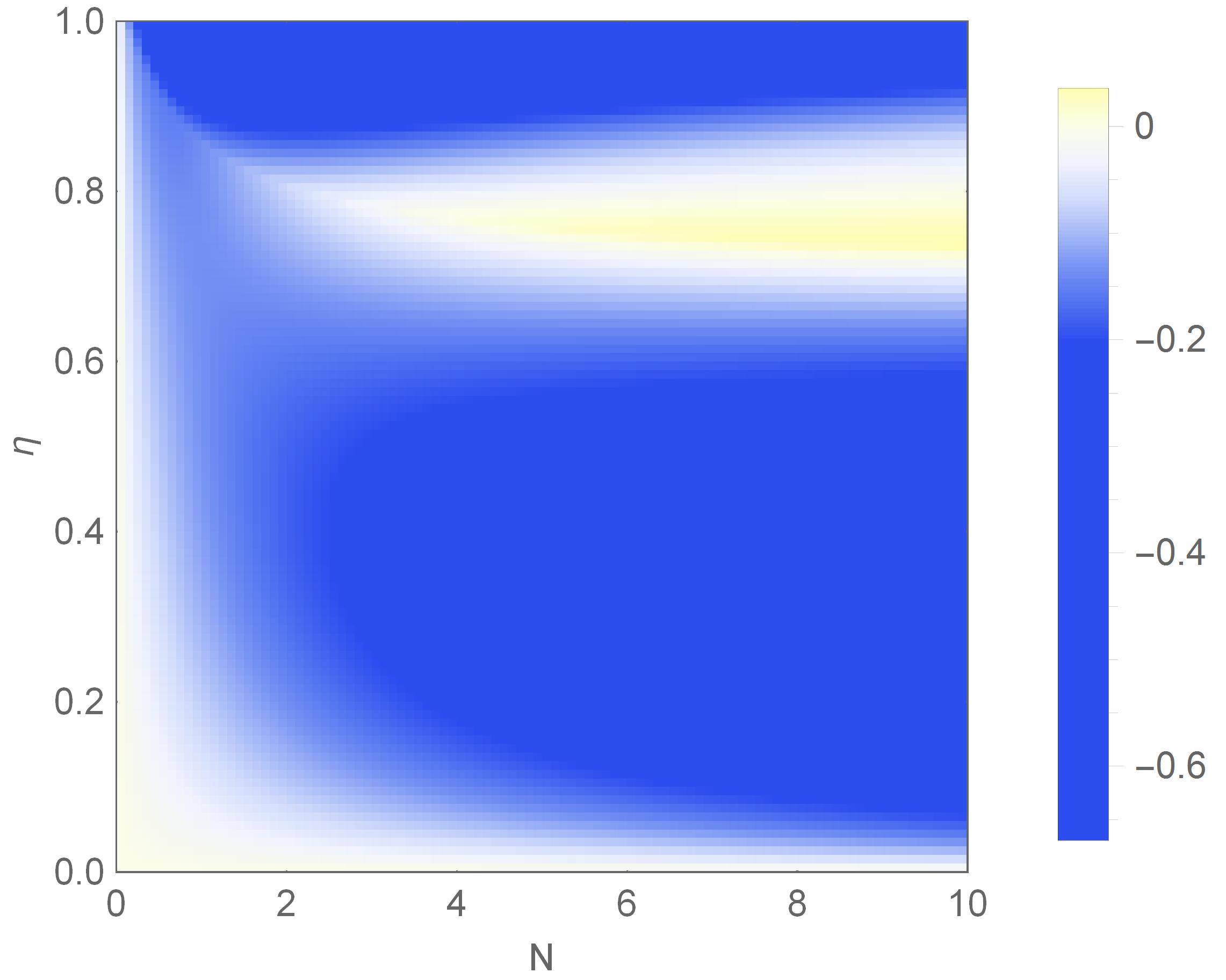}}
		\caption{(Color online) Relative increase $\mathcal I$ in precision for fixed $N$ obtained by using two-mode squeezed vacuum probes rather than the optimal single-mode input state, numerically evaluated for $10^4$ pairs $(N, \eta)\in [0,10]\times[0,1]$ and different values of $\epsilon$. \label{fig: increase vs N}}
	\end{figure}

In order to study the dependence upon $\epsilon$ of the results obtained in the noisy case, we can repeat the same numerical analysis performed in the main text for different $\epsilon$ values. In particular, for $\epsilon = 0.5$ and $0.1$, we plot in Figs. \ref{fig: optimal states}, \ref{fig: increase vs nA}, and \ref{fig: increase vs N} respectively the optimal single-mode displacement ratio $|\xi|^2/(2n_A)$, the increase in precision $\mathcal I$ for fixed $n_A$, and that for fixed total photon number $N$. We see that the same qualitative behavior is retained. However, the transition of the optimal displacement ratio $|\xi|^2/(2 n_A)$ from $0$ to $1$ is much faster when $\epsilon$ decreases, and the advantage $\mathcal I$ brought by correlated probes is less significant for smaller values of $\epsilon$.


\end{document}